\pdfoutput=1
\documentclass[11pt]{article}

\usepackage{jheppub}

\usepackage[english]{babel}

\usepackage{subfigure}
\usepackage{placeins}

\usepackage{braket}   
\usepackage{bbold}
\usepackage{appendix}

\newcommand{\sect}[1]{section~#1}

\newcommand{\fig}[1]{figure~#1}

\newcommand{\tab}[1]{table~#1}

\newcommand{\half}{{\textstyle\frac{1}{2}}}

\newcommand{\ms}{\mskip 1.5mu}

\newcommand{\tr}{\operatorname{tr}}

\newcommand{\tvec}[1]{\boldsymbol{#1}}

\newcommand{\mvec}[1]{\vec{\mskip 0.5mu #1}\mskip 1.5mu}

\newcommand{\Op}{\mathcal{O}}

\newcommand{\dd}{\mathrm{d}}
\newcommand{\pup}{p^{\ms\uparrow}}

\newcommand{\uup}{u^{\ms\uparrow}}
\newcommand{\udn}{u^{\ms\downarrow}}
\newcommand{\dup}{d^{\ms\uparrow}}
\newcommand{\ddn}{d^{\ms\downarrow}}

\newcommand{\ubup}{\bar{u}^{\ms\uparrow}}
\newcommand{\ubdn}{\bar{u}^{\ms\downarrow}}
\newcommand{\dbup}{\bar{d}^{\ms\uparrow}}
\newcommand{\dbdn}{\bar{d}^{\ms\downarrow}}

\newcommand{\avsum}{\sideset{}{^\prime}\sum}

\usepackage{tikz}
\graphicspath{{plots/},{graphs/}}

\allowdisplaybreaks[2]

\title{Double parton distributions with flavor interference from lattice QCD}

\abstract{
We study double parton distributions with flavor interference in the nucleon and compare them with previous results for the flavor diagonal case. We investigate both unpolarized and polarized partons.  We compare our lattice results with those obtained from the simple description of the proton in terms of an SU(6) symmetric three-quark wave function and find that this description fails for both flavor and polarization dependence.  We also derive and test a factorization ansatz for the unpolarized flavor interference distribution in terms of single-parton distributions and find that this ansatz fails to a large extent.
}

\preprint{\vbox{
\hbox{DESY-24-014}}}

\author[a]{Daniel Reitinger,}
\author[b]{Christian Zimmermann,}
\author[c]{Markus Diehl}
\author[a]{and Andreas Sch{\"a}fer}

\affiliation[a]{Institute for Theoretical Physics, University of Regensburg, 93040 Regensburg, Germany}
\affiliation[b]{Aix Marseille Univ, Université de Toulon, CNRS, CPT, Marseille, France}
\affiliation[c]{Deutsches Elektronen-Synchrotron DESY, Notkestr.~85, 22607 Hamburg, Germany}

\emailAdd{daniel.reitinger@ur.de}
\emailAdd{christian.zimmermann@univ-amu.fr}
\emailAdd{markus.diehl@desy.de}
\emailAdd{andreas.schaefer@physik.uni-regensburg.de}

\begin{document}

\maketitle

\section{Introduction}
\label{sec:intro}

The Standard Model of particle physics has been extremely successful in describing experimental data at high energies. As part of this, the fundamental QCD Lagrangian has been established beyond any reasonable doubt. However, many aspects of the resulting hadron properties are still only poorly understood, {e.g.}\ the entanglement of two partons in the wave functions of a proton. Work on this front is not only motivated by the goal to better understand the strong interaction, but also by the fact that QCD effects often limit the sensitivity of searches for physics beyond the Standard Model.

An intriguing phenomenon in hadron collisions is double parton scattering (DPS), a mechanism in which two partons in each hadron take part in a hard scattering subprocess. Building on pioneering work from the 1970s and 1980s \cite{Landshoff:1978fq, Kirschner:1979im, Politzer:1980me, Paver:1982yp, Shelest:1982dg, Mekhfi:1983az, Sjostrand:1986ep}, substantial progress has been made during the last decade in an effort to develop a systematic description of DPS in QCD \cite{Blok:2010ge, Gaunt:2011xd, Ryskin:2011kk, Blok:2011bu, Diehl:2011yj, Manohar:2012jr, Manohar:2012pe, Ryskin:2012qx, Gaunt:2012dd, Blok:2013bpa, Diehl:2017kgu, Cabouat:2019gtm, Cabouat:2020ssr}. Experimental investigations of DPS started in the 1980s \cite{AxialFieldSpectrometer:1986dfj} and were followed by a wealth of studies at the Tevatron and the LHC, see for instance  \cite{CDF:1993sbj,CDF:1997lmq,D0:2009apj,D0:2015dyx} and \cite{LHCb:2012aiv,ATLAS:2013aph,CMS:2013huw,ATLAS:2016rnd,LHCb:2015wvu,LHCb:2020jse,CMS:2022pio}.

A crucial input for computing DPS are double parton distributions (DPDs), which describe the joint distribution of two partons inside a hadron.  They quantify several types of two-particle correlations in the proton wave function and are not well known. It is natural to explore to which extent lattice QCD calculations can provide guidance in this context. In two previous publications \cite{Bali:2020mij,Bali:2021gel}, we presented lattice computations of two-current correlation functions that can be related with the Mellin moments of DPDs \cite{Diehl:2011yj}. This generalizes the well-known relation between single-current matrix elements and the Mellin moments of single-parton distributions, which has been extensively studied in the literature \cite{Lin:2017snn}. More recently, it has been proposed in \cite{Zhang:2023wea, Jaarsma:2023woo} to study the full functional dependence of DPDs on the lattice in the LaMET approach \cite{Ji:2013dva,Ji:2020ect}. It will be very interesting to see to which extent this can be done in practice.

One should bear in mind that neither of these lattice approaches is sensitive to parton momentum fractions smaller than, say, $10^{-2}$, which are responsible for much of the phase space where DPS is observed.  (The same holds for quark models, which have  been used extensively to compute DPDs \cite{Chang:2012nw, Rinaldi:2013vpa, Broniowski:2013xba,  Rinaldi:2014ddl, Broniowski:2016trx, Kasemets:2016nio, Rinaldi:2016jvu, Rinaldi:2016mlk, Rinaldi:2018zng, Courtoy:2019cxq, Broniowski:2019rmu, Broniowski:2020jwk}.) However, partons with larger momentum fractions at low resolution scales are the ``seeds'' of evolution and thus have an imprint on partons with smaller momentum fractions at high scales. At high scales, large parton momentum fractions are probed in the production of heavy particles and thus of interest in searches for new physics. Moreover, larger parton momentum fractions become relevant when the products of the two hard scatters in DPS have a large rapidity separation. This is an interesting kinematic region, where DPS is often appreciable compared to single hard scattering. Finally, DPDs at larger momentum fractions are of interest in their own right from the point of view of exploring hadron structure.

With this in mind, we complement in the present paper our previous study \cite{Bali:2021gel} of DPDs in the nucleon.  Specifically, we investigate flavor interference DPDs, which are characterized by different flavors for the quark (or antiquark) initiating a specific hard scattering in the amplitude and in its complex conjugate. Such distributions contribute for instance to the double Drell-Yan process.  They were introduced in \cite{Diehl:2011yj}, were it was also pointed out that they do not mix with gluons under evolution, such that at small momentum fractions one can expect them to be small compared with flavor diagonal DPDs. At moderate or large momentum fractions (which are relevant for
Mellin moments) there is however no argument that flavor interference should be suppressed.  We will investigate whether this is the case by comparing the corresponding two-current matrix elements computed on the lattice.  We will furthermore compare our lattice results (both for the flavor diagonal and the interference case) with the predictions obtained from an $SU(6)$ symmetric three-quark wave function of the proton.

Our paper is organized as follows. In section \ref{sec:theory}, we review the different quantities relevant to our study and explain how they are related to each other. Details of the lattice setup we use are given in \ref{sec:lattice}. Sections \ref{sec:results} and \ref{sec:factorization} contain the results of our calculations. In section \ref{sec:results}, the Mellin moments for different combinations of flavor and polarization are presented and compared with the $SU(6)$ predictions. In section \ref{sec:factorization} we explore to which extent DPDs can be factorized in terms of single-parton distributions.  We summarize our findings in section \ref{sec:conclusions}. 

\section{Theory background}
\label{sec:theory}
  
\subsection{Definitions and properties}  
  
\begin{figure}
\begin{center}
\includegraphics[scale=0.4]{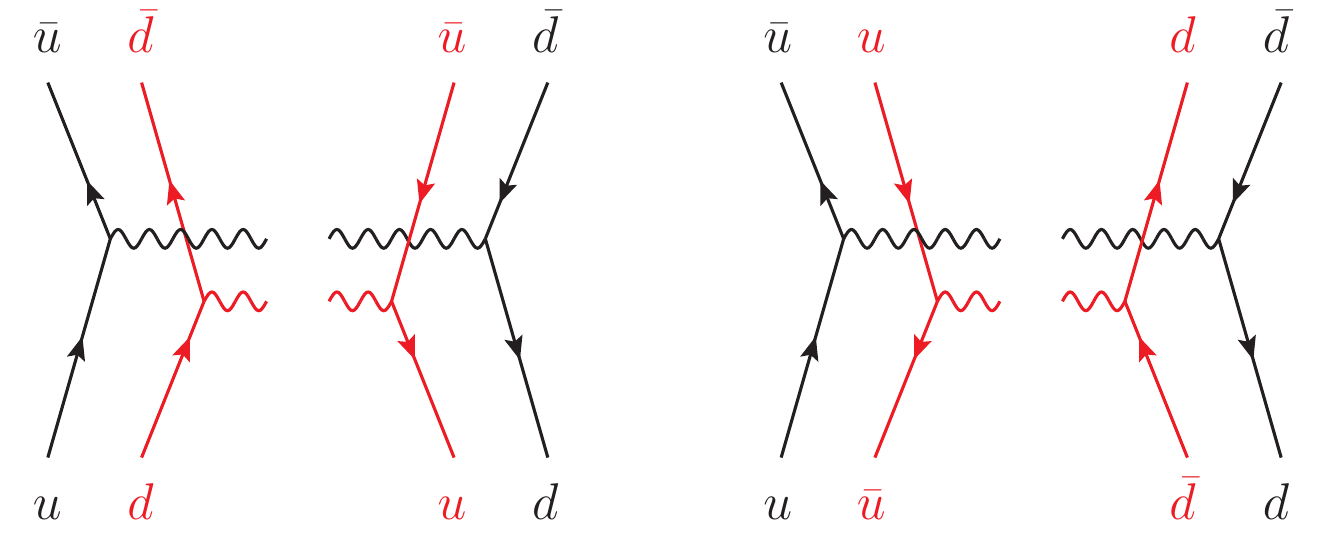}
\end{center}
\caption{Example graphs for flavor interference in the double Drell-Yan process.\label{fig:drell-yan-interf}}
\end{figure}

In the following, we review some basic definitions and properties of double parton distributions in the context of our lattice simulation. For more details the reader is referred to \cite{Bali:2020mij,Bali:2021gel}. DPDs describe the joint probability of finding two quarks with given polarization in a hadron. In this work we focus on the proton. We average over its polarization $\lambda$, which is indicated by the notation $\sum^\prime_\lambda = \frac{1}{2} \sum_\lambda$. The definition of DPDs is given by
  
\begin{align}
\label{eq:dpd-def}
F_{a_1 a_2}(x_1,x_2,\tvec{y})
= 2p^+ \int \dd y^- \int \frac{\dd z^-_1}{2\pi}\, \frac{\dd z^-_2}{2\pi}\,
&      e^{i\ms ( x_1^{} z_1^- + x_2^{} z_2^-)\ms p^+}
\nonumber \\
& \times
 \avsum_\lambda \bra{p,\lambda} \mathcal{O}_{a_1}(y,z_1)\, \mathcal{O}_{a_2}(0,z_2) \ket{p,\lambda}
\,,
\end{align}
where we use light-cone coordinates $v^\pm := (v^0 \pm v^3)/\sqrt{2}$ and $\tvec{v} := (v^1,v^2)$ for a given four-vector $v^\mu$. The light-cone operators are defined as

\begin{align}
\label{eq:quark-ops}
\mathcal{O}_{a}(y,z)
&= \bar{q}\bigl( y - \half z \bigr)\, \Gamma_{a} \, q\bigl( y + \half z \bigr)
   \Big|_{z^+ = y^+_{} = 0,\, \tvec{z} = \tvec{0}}\,,
\end{align}
where $a$ specifies the quark flavor and polarization, which is determined by the spin projections

\begin{align}
  \label{eq:quark-proj}
\Gamma_q & = \half \gamma^+ \,, &
\Gamma_{\Delta q} &= \half \gamma^+\gamma_5 \,, &
\Gamma_{\delta q}^j = \half i \sigma^{j+}_{} \gamma_5  \quad (j=1,2) \,.
\end{align}  
$q$ refers to an unpolarized quark, $\Delta q$ to a longitudinally polarized quark and $\delta q$ to a transversely polarized quark. The DPDs \eqref{eq:dpd-def} can be decomposed in terms of rotationally invariant functions:

\begin{align} 
\label{eq:invar-dpds}
F_{q_1 q_2}(x_1,x_2, \tvec{y}) &= f_{q_1 q_2}(x_1,x_2, y^2) \,,
\nonumber \\
F_{\Delta q_1 \Delta q_2}(x_1,x_2, \tvec{y})
&= f_{\Delta q_1 \Delta q_2}(x_1,x_2, y^2) \,,
\nonumber \\
F_{\delta q_1 q_2}^{j_1}(x_1,x_2, \tvec{y})
&= \epsilon^{j_1 k} \tvec{y}^k\, m f_{\delta q_1 q_2}(x_1,x_2, y^2) \,,
\nonumber \\
F_{q_1 \delta q_2}^{j_2}(x_1,x_2, \tvec{y})
&= \epsilon^{j_2 k} \tvec{y}^k\, m f_{q_1 \delta q_2}(x_1,x_2, y^2) \,,
\nonumber \\
F_{\delta q_1 \delta q_2}^{j_1 j_2}(x_1,x_2, \tvec{y})
&= \delta^{j_1 j_2} f^{}_{\delta q_1 \delta q_2}(x_1,x_2, y^2)
\nonumber \\
&\quad  + \bigl( 2 \tvec{y}^{j_1} \tvec{y}^{j_2}
         - \delta^{j_1 j_2} \tvec{y}^2 \bigr)\ms
   m^2 f^{\ms t}_{\delta q_1 \delta q_2}(x_1,x_2, y^2) \,,
\end{align}
where $m$ denotes the proton mass and $\epsilon^{jk}$ is the antisymmetric tensor in two dimensions ($\epsilon^{12}~=~1$). The definitions given above can be extended for the case of flavor-changing operators, i.e.\ \eqref{eq:quark-ops} is modified so that the quark field has a different quark flavor than the conjugate quark field, $q \neq q^\prime$:

\begin{align}
\label{eq:quark-ops-interf}
\mathcal{O}_{a}(y,z)
&= \bar{q}\bigl( y - \half z \bigr)\, \Gamma_{a} \, q^\prime\bigl( y + \half z \bigr)
   \Big|_{z^+ = y^+_{} = 0,\, \tvec{z} = \tvec{0}}\,,
\end{align}
with $a = (qq^\prime), \Delta(qq^\prime), \delta(qq^\prime)$. The corresponding functions $F_{a_1 a_2}(x_1,x_2,\tvec{y})$ given by inserting the operators \eqref{eq:quark-ops-interf} in \eqref{eq:dpd-def} are called flavor interference distributions or flavor interference DPDs. For instance, an interference contribution to the unpolarized channel is given by $F_{(ud)(du)}$, which corresponds to the operator combination $\mathcal{O}_{(ud)} \mathcal{O}_{(du)}$. Notice that in contrast to flavor diagonal DPDs, these interference DPDs cannot be interpreted as parton density distributions. Hence, there is no positivity constraint for them. Flavor interference DPDs are relevant in the description of double parton scattering, where they represent flavor interference contributions in the cross section. An example is shown in \fig\ref{fig:drell-yan-interf}.

\begin{figure}
\begin{center}
\includegraphics[scale=0.54]{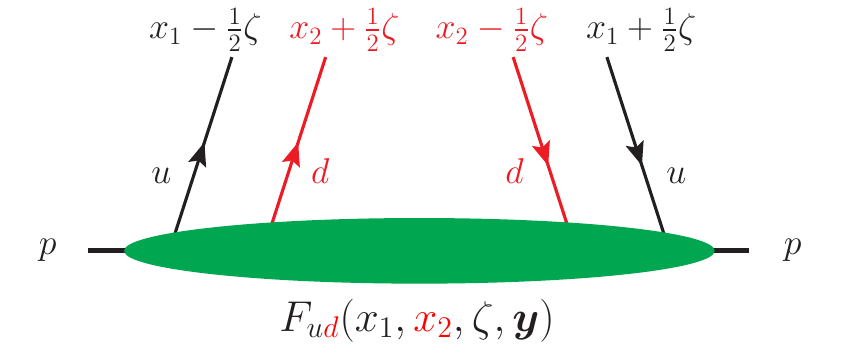}
\includegraphics[scale=0.54]{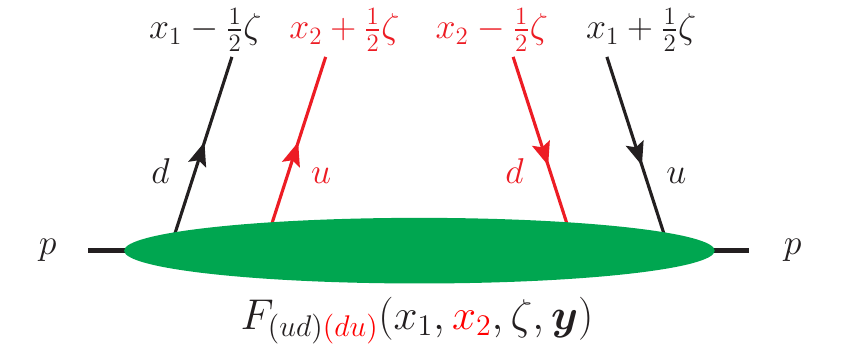}
\end{center}
\caption{Skewed DPDs and their parton momentum fractions for the flavor
diagonal case (left) and for flavor interference (right).\label{fig:dpd-kinem}}
\end{figure}

In the context of lattice calculations, it is useful to introduce so-called skewed DPDs, which are a generalization of ordinary DPDs in the sense that one introduces a difference between the momentum fractions of the quark in the wave function and in its complex conjugate. This difference is quantified by the skewness parameter $\zeta$: 

\begin{align}
\label{eq:dpd-skew-def}
F_{a_1 a_2}(x_1,x_2,\zeta, \tvec{y})
&= 2p^+ \int \dd y^- e^{-i \zeta y^- p^+}
  \int \frac{\dd z^-_1}{2\pi}\, \frac{\dd z^-_2}{2\pi}\,
          e^{i\ms ( x_1^{} z_1^- + x_2^{} z_2^-)\ms p^+}
\nonumber \\
& \qquad \times
    \avsum_\lambda \bra{p,\lambda} \mathcal{O}_{a_1}(y,z_1)\, \mathcal{O}_{a_2}(0,z_2)
    \ket{p,\lambda} \,.
\end{align}
This skewed DPD has already been used in \cite{Bali:2020mij,Bali:2021gel}, where also the region of support w.r.t.\ the parameters $x_1$, $x_2$ and $\zeta$ has been discussed in detail.  Let us recall that this support region is given by:

\begin{align}
\label{eq:support-region}
|x_i \pm \half \zeta| \le 1\,, \qquad |x_1|+|x_2| \le 1\,, \qquad |\zeta| \le 1\,.
\end{align}  
In \fig\ref{fig:dpd-kinem} we give a graphical representation of skewed DPDs, where we indicate the longitudinal momentum fraction of the quarks in the wave function and its complex conjugate.

Moreover, we define Mellin moments w.r.t.\ $x_1$ and $x_2$:

\begin{align}
  \label{eq:skewed-inv-mellin-mom-def}
I_{a_1 a_2}(\zeta, y^2)
&= \int_{-1}^{1} \dd x_1^{} \int_{-1}^{1} \dd x_2^{} \;
   f_{a_1 a_2}(x_1,x_2,\zeta, y^2) \,.
\end{align}  
The symmetry properties of flavor diagonal DPDs are discussed in \cite{Bali:2021gel}. In the following we briefly show how these properties generalize to the flavor interference case.

For unpolarized distributions, one finds
\begin{align}
   \label{unpol-sym-rel}
   f_{(ud) (du)}(x_1, x_2, \zeta, y^2)
   &= f_{(du) (ud)}(x_1, x_2, - \zeta, y^2)
   \,,
   \notag \\
   \bigl[ f_{(ud) (du)}(x_1, x_2, \zeta, y^2) \bigr]^*
   &= f_{(du) (ud)}(x_1, x_2, - \zeta, y^2)
   \,,
   \notag \\
   f_{(ud) (du)}(x_1, x_2, \zeta, y^2)
   &= f_{(du) (ud)}(x_2, x_1, - \zeta, y^2)
   \,,
\end{align}
where the first relation follows from $PT$ invariance, the second one from taking the hermitian conjugate of the definition \eqref{eq:dpd-skew-def}, and the third one from interchanging the two operators.  From the first two relations, it follows that $f_{(ud) (du)}$ is real valued.  Equations analogous to \eqref{unpol-sym-rel} relate $f_{\Delta (ud) \ms \Delta (du)}$ with $f_{\Delta (du) \ms \Delta (ud)}$,  $f_{\delta (ud) \ms \delta (du)}$ with $f_{\delta (du) \ms \delta (ud)}$,  and $f^{t}_{\delta (ud) \ms \delta (du)}$ with $f^{t}_{\delta (du) \ms \delta (ud)}$. For a single transverse polarization, we have instead
\begin{align}
   \label{transv-sym-rel}
   f_{(ud) \ms \delta (du)}(x_1, x_2, \zeta, y^2)
   &= f_{(du) \ms \delta (ud)}(x_1, x_2, - \zeta, y^2)
   \,,
   \notag \\
   \bigl[ f_{(ud) \ms \delta (du)}(x_1, x_2, \zeta, y^2) \bigr]^*
   &= f_{(du) \ms \delta (ud)}(x_1, x_2, - \zeta, y^2)
   \,,
   \notag \\
   f_{(ud) \ms \delta (du)}(x_1, x_2, \zeta, y^2)
   &= {}- f_{\delta (du) (ud)}(x_2, x_1, - \zeta, y^2)
   \,.
\end{align}
Also in this case, we find that the distributions are real valued.  Parity invariance implies that $F_{(ud) \ms \Delta (du)}$, $F_{(du) \ms \Delta (ud)}$, $F_{\Delta (ud) (du)}$, and $F_{\Delta (du) (ud)}$ are zero.

\vspace*{1cm} 

\subsection{Euclidean matrix elements}

Information about DPDs can be obtained from first principles on the lattice through Euclidean two-current matrix elements. This has been worked out in detail for the pion \cite{Bali:2020mij} and the nucleon \cite{Bali:2021gel}. In the following, we recall the important definitions and relations to DPDs and extend them to the case of flavor interference. The Euclidean two-current matrix element of the nucleon is defined as:
    
\begin{align}
  \label{eq:mat-els}
M^{\mu_1 \cdots \mu_2 \cdots}_{q_1 q_2 q_3 q_4, i_1 i_2}(p,y)
&:= \avsum_\lambda \bra{p,\lambda} J^{\mu_1 \cdots}_{q_1 q_2, i_1}(y)\,
              J^{\mu_2 \cdots}_{q_3 q_4, i_2}(0) \ket{p,\lambda} \,,
\end{align}
where we take the average of the proton spin. The currents $J^{\mu\dots}_{qq^\prime,i}$ are local quark bilinear operators. In this work we focus on three types of currents, which are defined as:

\begin{align}
  \label{eq:local-ops}
J_{qq^\prime, V}^\mu(y) &= \bar{q}(y) \ms \gamma^\mu\ms q^\prime(y) \,,
&
J_{qq^\prime, A}^\mu(y) &= \bar{q}(y) \ms \gamma^\mu \gamma_5\, q^\prime(y) \,,
&
J_{qq^\prime, T}^{\mu\nu}(y) &= \bar{q}(y) \ms \sigma^{\mu\nu} \ms q^\prime(y) \,.
\end{align}    
In order to make contact with the DPDs, we decompose the matrix elements in terms of Lorentz invariant functions:

\begin{align}
  \label{eq:tensor-decomp}
M^{\{\mu\nu\}}_{q_1 q_2 q_3 q_4, V V}
        - \tfrac{1}{4} g^{\mu\nu} g_{\alpha\beta}
                M^{\alpha\beta}_{q_1 q_2 q_3 q_4, V V}
 & = u_{V V,A}^{\mu\nu}\, A_{(q_1 q_2) (q_3 q_4)}^{}
   + u_{V V,B}^{\mu\nu}\, m^2\ms B_{(q_1 q_2) (q_3 q_4)}^{}
\nonumber \\[0.1em]
 &\quad + u_{V V,C}^{\mu\nu}\, m^4\ms C_{(q_1 q_2) (q_3 q_4)}^{} \,,
\nonumber \\[0.1em]
M^{\mu\nu\rho}_{q_1 q_2 q_3 q_4, T V} + \tfrac{2}{3} g_{\vphantom{q_1T}}^{\rho[\mu} M^{\nu]\alpha\beta}_{q_1 q_2 q_3 q_4, T V} g_{\alpha\beta}
 & = u_{T V,A}^{\mu\nu\rho}\, m\ms A_{\delta (q_1 q_2) (q_3 q_4)}^{}
   + u_{T V,B}^{\mu\nu\rho}\,  m^3\ms B_{\delta (q_1 q_2) (q_3 q_4)}^{} \,,
\nonumber \\[0.3em]
\tfrac{1}{2} \, \bigl[ M^{\mu\nu\rho\sigma}_{q_1 q_2 q_3 q_4, T T}
     + M^{\rho\sigma\mu\nu}_{q_1 q_2 q_3 q_4, T T} \bigr]
 &= \tilde{u}_{T T,A}^{\mu\nu\rho\sigma}\, A_{\delta (q_1 q_2) \delta (q_3 q_4)}^{}
    + \tilde{u}_{T T,B}^{\mu\nu\rho\sigma}\, m^2\ms B_{\delta (q_1 q_2) \delta (q_3 q_4)}^{}
\nonumber \\[0.3em]
 &\quad 	+ \tilde{u}_{T T,C}^{\mu\nu\rho\sigma}\, m^2\ms C_{\delta (q_1 q_2) \delta (q_3 q_4)}^{}
    + \tilde{u}_{T T,D}^{\mu\nu\rho\sigma}\, m^4\ms D_{\delta (q_1 q_2) \delta (q_3 q_4)}^{}
\nonumber \\[0.3em]
 &\quad   + u_{T T,E}^{\mu\nu\rho\sigma}\, m^2\ms \widetilde{E}_{\delta (q_1 q_2) \delta (q_3 q_4)}^{} \,,
\end{align}
where the basis tensors $u$ and $\tilde{u}$, which depend on the Lorentz vectors $y$ and $p$, have been defined in \cite{Bali:2021gel}, equation (2.26). The quantities $A_{(q_1 q_2)(q_3 q_4)}$, etc. are Lorentz scalar functions depending only on $y^2$ and $py$. At leading twist, we only need to consider the functions $A_{(q_1 q_2)(q_3 q_4)}$, $A_{\Delta(q_1 q_2)\Delta(q_3 q_4)}$, $A_{\delta(q_1 q_2)(q_3 q_4)}$, $A_{(q_1 q_2)\delta(q_3 q_4)}$, $A_{\delta(q_1 q_2)\delta(q_3 q_4)}$, and $B_{\delta(q_1 q_2)\delta(q_3 q_4)}$. These so-called twist-two functions are directly related to the Mellin moments \eqref{eq:skewed-inv-mellin-mom-def}:

\begin{align}
\label{eq:skewed-mellin-inv-fct}
I_{a_1 a_2}(\zeta, y^2)
&= \int_{-\infty}^{\infty} \dd(py)\, e^{-i\zeta py}\, A_{a_1 a_2}(py,y^2) \,,
\\
\label{eq:skewed-mellin-inv-fct-quad}
I^t_{a_1 a_2}(\zeta, y^2)
&= \int_{-\infty}^{\infty} \dd(py)\, e^{-i\zeta py}\, B_{a_1 a_2}(py,y^2) \,,
\end{align}
where \eqref{eq:skewed-mellin-inv-fct-quad} is only defined for transverse polarization of $a_1$ and $a_2$. In this work, we shall restrict ourselves to $\mvec{p}=\mvec{0}$ and, therefore, $py = 0$. In this case, the twist-two functions correspond to the first moment in $\zeta$ of the DPD:

\begin{align}
\label{eq:py-zero-fct}
A_{a_1 a_2}(py=0, y^2) = \frac{1}{2\pi}
  \int_{-1}^1 \dd \zeta \, I_{a_1 a_2}(\zeta, y^2) \,.
\end{align}
In \cite{Bali:2021gel} we found similar patterns for the twist-two functions and the reconstructed DPDs themselves regarding their dependence on the distance $y$ and on the quark polarization.
  
\subsection{DPDs in the $SU(6)$ quark model}  
\label{sec:dpd-su6}

In the following, we consider a simple $SU(6)$-symmetric quark model and derive its predictions for matrix elements of two-quark currents. The spin-flavor part of the $SU(6)$-symmetric proton wave function $\ket{\pup}$ is given by:

\begin{align}
\ket{\pup} &= \frac{1}{3\sqrt{2}} 
\left[ 
\ket{\uup\udn\dup} + \ket{\udn\uup\dup} - 2\ket{\uup\uup\ddn} +
\ket{\uup\dup\udn} + \ket{\udn\dup\uup} - 2\ket{\uup\ddn\uup} +
\right.\nonumber \\ &\quad \left. +
\ket{\dup\uup\udn} + \ket{\dup\udn\uup} - 2\ket{\ddn\uup\uup} 
\right] \,,
\label{eq:proton-state}
\end{align}
where $\uparrow$ ($\downarrow$) indicates polarization along the positive (negative) $z$ axis. Moreover, the quark operators are expressed as

\begin{align}
\Op_{(ud)} &= \frac{1}{2} \left[ (\ubup\gamma^+\dup) + (\ubdn\gamma^+\ddn) \right] \,, \nonumber \\
\Op_{\Delta (ud)} &= \frac{1}{2} \left[ (\ubup\gamma^+\dup) - (\ubdn\gamma^+\ddn) \right] \,, 
\label{eq:su6-quark-ops}
\end{align}
and likewise for the other flavor combinations. Considering matrix elements of the form $\bra{\pup} (\bar{q}_1 \gamma^+ q_2) (\bar{q}_3 \gamma^+ q_4) \ket{\pup}$ for the proton state \eqref{eq:proton-state}, we obtain:

\begin{align}
\bra{\pup} (\ubup\gamma^+\uup)(\dbup\gamma^+\dup) \ket{\pup} &= a \,, &\quad
\bra{\pup} (\ubup\gamma^+\uup)(\dbdn\gamma^+\ddn) \ket{\pup} &= 4a \,, \nonumber \\
\bra{\pup} (\ubdn\gamma^+\udn)(\dbup\gamma^+\dup) \ket{\pup} &= a \,, &\quad
\bra{\pup} (\ubdn\gamma^+\udn)(\dbdn\gamma^+\ddn) \ket{\pup} &= 0 \,, \nonumber \\
\bra{\pup} (\ubup\gamma^+\uup)(\ubup\gamma^+\uup) \ket{\pup} &= 4a \,, &\quad
\bra{\pup} (\ubup\gamma^+\uup)(\ubdn\gamma^+\udn) \ket{\pup} &= a \,, \nonumber \\
\bra{\pup} (\ubdn\gamma^+\udn)(\ubup\gamma^+\uup) \ket{\pup} &= a \,, &\quad
\bra{\pup} (\ubdn\gamma^+\udn)(\ubdn\gamma^+\udn) \ket{\pup} &= 0 \,, \nonumber \\
\bra{\pup} (\dbup\gamma^+\uup)(\ubup\gamma^+\dup) \ket{\pup} &= a \,, &\quad
\bra{\pup} (\dbup\gamma^+\uup)(\ubdn\gamma^+\ddn) \ket{\pup} &= -2a \,, \nonumber \\
\bra{\pup} (\dbdn\gamma^+\udn)(\ubup\gamma^+\dup) \ket{\pup} &= -2a \,, &\quad
\bra{\pup} (\dbdn\gamma^+\udn)(\ubdn\gamma^+\ddn) \ket{\pup} &= 0 \,,
\label{eq:su6-me-parts}
\end{align}
where it is understood that the four field operators are taken at different positions as specified by \eqref{eq:dpd-def} and \eqref{eq:quark-ops}. Here the factor $a$ depends on the orbital part of the proton wave function, which we leave unspecified. The DPDs are expressed in terms of proton matrix elements of the operators \eqref{eq:su6-quark-ops}, i.e.\ $f_{a_1 a_2} \propto \bra{\pup} \Op_{a_1} \Op_{a_2} \ket{\pup}$. Using the results given in \eqref{eq:su6-me-parts}, we find:

\begin{align}
f_{ud} &= +6\tilde{a}\,, &\quad 
f_{uu} &= +6\tilde{a}\,, &\quad 
f_{(du)(ud)} &= -3\tilde{a} \,, \nonumber \\
f_{\Delta u \Delta d} &= -4\tilde{a} \,, &\quad 
f_{\Delta u \Delta u} &= +2\tilde{a} \,, &\quad 
f_{\Delta (du) \Delta (ud)} &= +5\tilde{a} \,.
\end{align}
with an overall factor $\tilde{a}$ that depends again on the orbital part of the wave function. Since this orbital part is isotropic, there is no difference regarding the direction of the quark polarization. Hence, the values obtained for $f_{\Delta u \Delta d}$ are the same for $f_{\delta u \delta d}$, and likewise for all other flavor combinations. Predictions that are independent of the factor $\tilde{a}$ are obtained for \emph{ratios} of DPDs, such as:

\begin{align}
\frac{f_{(du)(ud)}}{f_{ud}} = -\frac{1}{2}\,, \qquad\quad
\frac{f_{(du)(ud)}}{f_{uu}} = -\frac{1}{2}\,, \qquad\quad
\frac{f_{ud}}{f_{uu}} = +1\,,
\label{eq:su6-ratios}
\end{align}
and

\begin{align}
\frac{f_{\Delta(du)\Delta(ud)}}{f_{ud}} = +\frac{5}{6} \,, \qquad\quad
\frac{f_{\Delta u \Delta d}}{f_{ud}} = -\frac{2}{3}\,, \qquad\quad
\frac{f_{\Delta u \Delta u}}{f_{uu}} = +\frac{1}{3}\,.
\label{eq:su6-ratios-pol}
\end{align}
Analogous predictions hold for the ratios of the twist-two functions $A_{a_1 a_2}$ and can be directly checked against lattice results.

\section{Lattice calculation}
\label{sec:lattice}

In the following, we give an overview of the calculation of two-current matrix elements on the lattice and briefly review the techniques that are used. A detailed explanation is given in \cite{Bali:2021gel}. This shall now be extended to flavor-changing operators.

Let us first recall that the two-current matrix element \eqref{eq:mat-els} of the nucleon at $y^0 = 0$ is related to the nucleon four-point function $C_{4\mathrm{pt}}(\mvec{y},t,\tau)$ by the following formula:

\begin{align}
\left. M_{ij}(p,y) \vphantom{\sum} \right|_{y^0 = 0} = C^{ij,\mvec{p}}_{\mathrm{4pt}}(\mvec{y}) := 
	2V \sqrt{m^2 + \mvec{p}^2} 
	\left. 
		\frac{C^{ij,\mvec{p}}_{\mathrm{4pt}}(\mvec{y},t,\tau)}
		{C^{\mvec{p}}_\mathrm{2pt}(t)} 
	\right|_{0 \ll \tau \ll t} \,,
\label{eq:4pt-2pt-ratio}
\end{align}
where $V$ is the spatial lattice volume and the four-point function $C^{ij,\mvec{p}}_{\mathrm{4pt}}(\mvec{y},t,\tau)$ is given by:

\begin{align}
C^{ij,\mvec{p}}_{\mathrm{4pt}}(\mvec{y},t,\tau) 
&:= 
	a^6 \sum_{\mvec{z}^\prime,\mvec{z}} 
	e^{-i\mvec{p}(\mvec{z}^\prime-\mvec{z})}\  
	\left\langle \tr \left\{
		P_+ \mathcal{P}(\mvec{z}^\prime,t)\ J_i(\mvec{y},\tau)\ 
		J_j(\mvec{0},\tau)\ \overline{\mathcal{P}}(\mvec{z},0) 
	\right\} \right\rangle\,. 
\label{eq:4ptdef}
\end{align}
Here $P_+ = \left( \mathbb{1} + \gamma_4 \right)/2$ projects onto positive parity, and $\overline{\mathcal{P}}(\mvec{x},t)$ and $\mathcal{P}(\mvec{x},t)$ are the nucleon interpolators, for which we choose:

\begin{align}
\overline{\mathcal{P}}(\mvec{x},t) &:= 
	\left.\epsilon_{abc}\ 
		\left[ 
			\bar{u}_a(x)\ C \gamma_5\ \bar{d}_b^{\,T}(x) 
		\right] \bar{u}_c(x) 
	\vphantom{\sum}\right|_{x^4=t} \,, 
	\nonumber\\
\mathcal{P}(\mvec{x},t) &:= 
	\left.\epsilon_{abc}\ u_a(x) 
	\left[ 
		u_b^T(x)\ C \gamma_5\ d_c(x) 
	\right] \vphantom{\sum}\right|_{x^4=t} \,,
\label{eq:interpdef}
\end{align}
where $C$ is the charge conjugation matrix in Dirac space. The two-point function $C^{\mvec{p}}_{\mathrm{2pt}}(t)$ appearing in \eqref{eq:4pt-2pt-ratio} is defined as:

\begin{align}
C^{\mvec{p}}_{\mathrm{2pt}}(t) &:= 
	a^6 \sum_{\mvec{z}^\prime,\mvec{z}} 
	e^{-i\mvec{p}(\mvec{z}^\prime-\mvec{z})}\  
	\left\langle \tr \left\{ 
		P_+ \mathcal{P}(\mvec{z}^\prime,t)\ 
		\overline{\mathcal{P}}(\mvec{z},0) \right\} 
	\right\rangle\,.
\label{eq:2ptdef}
\end{align}
\paragraph{Wick contractions:}

\begin{figure}
\begin{center}
\includegraphics[scale=1]{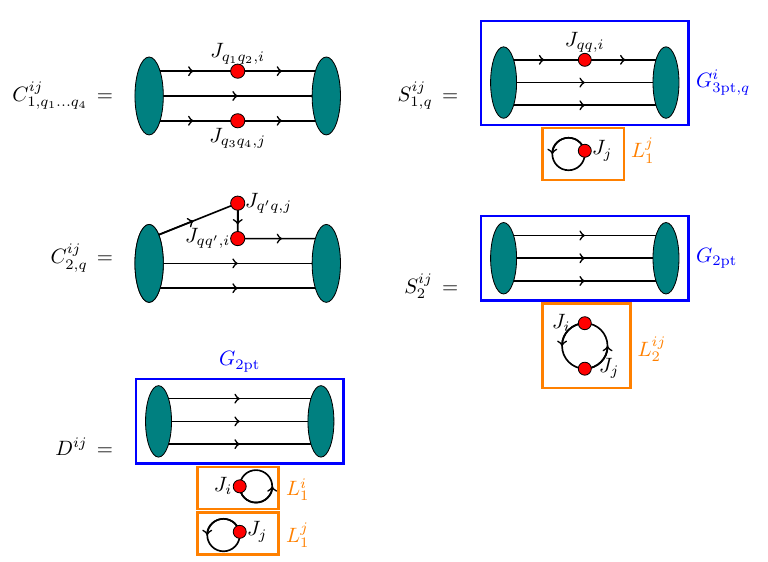}
\end{center}
\caption{Depiction of the five types of Wick contractions that contribute to a nucleon four-point function. In case of $C_1$, $C_2$ and $S_1$ the explicit contraction depends on all involved flavors. Provided that the quark masses are taken to be equal, $C_2$ depends only on the flavor of the quark line that is connected to one of the insertions (red) and the source. For the disconnected diagrams $S_1$, $S_2$ and $D$ we also indicate the disconnected parts $G_{3\mathrm{pt}}$ and $G_{2\mathrm{pt}}$ (blue), as well as the loops $L_1$ and $L_2$ (orange).\label{fig:graphs}}
\end{figure}

The four-point function \eqref{eq:4ptdef} decomposes into a definite set of Wick contractions w.r.t.\ the fermion fields. There are five types of contractions, which we call $C_1$, $C_2$, $S_1$, $S_2$ and $D$. These contractions are represented by the graphs in \fig\ref{fig:graphs}. Notice that, depending on the quark flavor of the operators, there are several contributions for each contraction type. For $C_1$-type graphs, we denote this by the flavor indices of the operator insertions $J_{q_1 q_2,i}$ and $J_{q_3 q_4,j}$, i.e.\ $C_{1,q_1 q_2 q_3 q_4}$. For the proton, only the contractions $C_{1,uudd}$, $C_{1,uuuu}$ and $C_{1,duud}$ contribute (together with $C_{1,dduu}$ and $C_{1,uddu}$, which are obtained by exchanging the two currents). In our calculation, we consider only proton matrix elements in the iso-symmetric limit, where the $u$-quark and the $d$-quark have the same mass. In that case, there are two independent contributions with $C_2$- or $S_1$-topology, namely $C_{2,u}$, $C_{2,d}$, $S_{1,u}$ and $S_{1,d}$. The indicated flavor refers to the quark line that connects one of the currents with the proton source.

The explicit set of contractions contributing to the four-point function depends on the quark flavors of the considered operators. In the following we list the decomposition of all possible matrix elements of two light-quark operators in a proton:
\begin{align}
\left. M_{uudd, ij}(p,y)\right|_{y^0 = 0} 
&= 
	C^{ij,\mvec{p}}_{1,uudd}(\mvec{y}) + 
	S^{ij,\mvec{p}}_{1,u}(\mvec{y}) + 
	S^{ji,\mvec{p}}_{1,d}(-\mvec{y}) + 
	D^{ij,\mvec{p}}(\mvec{y})\,,
\nonumber \\
\left. M_{uuuu, ij}(p,y)\right|_{y^0 = 0} 
&= 
	C^{ij,\mvec{p}}_{1,uuuu}(\mvec{y}) + 
	C^{ij,\mvec{p}}_{2,u}(\mvec{y}) + 
	C^{ji,\mvec{p}}_{2,u}(-\mvec{y}) + 
	S^{ij,\mvec{p}}_{1,u}(\mvec{y}) + 
	S^{ji,\mvec{p}}_{1,u}(-\mvec{y})
\nonumber \\
&\quad + 
	S_{2}^{ij,\mvec{p}}(\mvec{y}) + 
	D^{ij,\mvec{p}}(\mvec{y})\,,
\nonumber \\
\left. M_{dddd, ij}(p,y)\right|_{y^0 = 0} 
&= 
	C^{ij,\mvec{p}}_{2,d}(\mvec{y}) + 
	C^{ji,\mvec{p}}_{2,d}(-\mvec{y}) + 
	S^{ij,\mvec{p}}_{1,d}(\mvec{y}) + 
	S^{ji,\mvec{p}}_{1,d}(-\mvec{y}) 
\nonumber \\
&\quad +
	S_{2}^{ij,\mvec{p}}(\mvec{y}) + 
	D^{ij,\mvec{p}}(\mvec{y})\,,
	\nonumber \\
\left. M_{duud, ij}(p,y)\right|_{y^0 = 0} 
&= 
	C^{ij,\mvec{p}}_{1,duud}(\mvec{y}) + 
	C^{ij,\mvec{p}}_{2,d}(\mvec{y}) + 
	C^{ji,\mvec{p}}_{2,u}(-\mvec{y}) + 
	S^{ij,\mvec{p}}_{2}(\mvec{y})\,,
	\nonumber \\
\left. M_{uddu, ij}(p,y)\right|_{y^0 = 0} 
&= 
	C^{ji,\mvec{p}}_{1,duud}(-\mvec{y}) + 
	C^{ij,\mvec{p}}_{2,u}(\mvec{y}) + 
	C^{ji,\mvec{p}}_{2,d}(-\mvec{y}) + 
	S^{ij,\mvec{p}}_{2}(\mvec{y})\,,
\label{eq:phys_me_decomp}
\end{align}
where we use

\begin{align}
C^{ij,\mvec{p}}_{1,uudd}(\mvec{y}) = 
	2V \sqrt{m^2+\mvec{p}^2} 
	\left. 
		\frac{
			C^{ij,\mvec{p}}_{1,uudd}(\mvec{y},t,\tau)
		}{
			C_{2\mathrm{pt}}^{\mvec{p}}(t)
		} 
	\right|_{0 \ll \tau \ll t}
\end{align}
and similarly for all other Wick contractions.

\paragraph{Renormalization:}

The lattice operators are renormalized multiplicatively and converted to the $\overline{\mathrm{MS}}$-scheme using the factors $Z_i$:

\begin{align}
\label{eq:latt_op_ren}
J_i^{\overline{\mathrm{MS}}}(y) = Z_i J_i^{\mathrm{latt}}(y)\,.
\end{align}
For $\beta = 3.4$, the corresponding values of $Z_i$ are \cite{RQCD:2020kuu}:

\begin{align}
Z_V = 0.7128\,, \qquad Z_A = 0.7525\,, \qquad Z_T = 0.8335\,,
\end{align}
for the choice

\begin{align}
\label{eq:the_scale}
\mu = 2~\mathrm{GeV}\,.
\end{align}
The renormalization of the two-current matrix elements is given by:

\begin{align}
M^{\overline{\mathrm{MS}}}_{q_1 q_2 q_3 q_4, i_1 i_2} = 
Z_{i_1} Z_{i_2} M^{\mathrm{latt}}_{q_1 q_2 q_3 q_4, i_1 i_2}\,.
\end{align}
%

\paragraph{Technical details on Wick contractions:}

\begin{figure}[ht]
\includegraphics[scale=1]{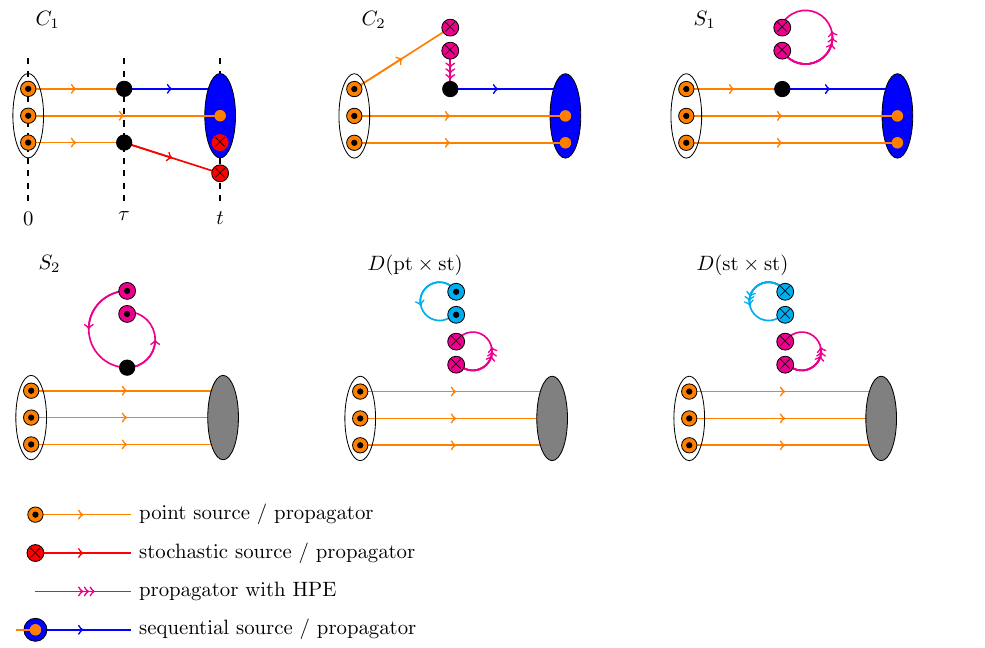}
\caption{Sketch of all Wick contractions that are considered in this work. Colors have no meaning regarding the evaluation technique but only indicate to which source the corresponding propagator belongs. We use two evaluation methods for the $D$ graph, which are drawn at the bottom center and the bottom right. \label{fig:graphs_tech}}
\end{figure}

A sketch summarizing the techniques used to evaluate the Wick contractions is given in \fig\ref{fig:graphs_tech}. The proton source is realized by a point source. Both the proton source and the proton sink are momentum smeared \cite{Bali:2016lva}. The corresponding smeared point-to-all propagator at the source $z$ is denoted by $M_z^{\Phi,\mvec{p}}(y)$. The contractions where at least one of the two currents is directly connected to the proton source or sink require usage of the sequential source technique. $C_1$ and $C_2$ additionally involve stochastic wall sources for a given timeslice $t$, for which we use $Z_2 \times Z_2$ sources. We denote the corresponding propagated stochastic source ("stochastic propagator") by $\psi_t^{(\ell)}$, where $\ell$ indicates the stochastic source. In the case of $C_2$, the stochastic source timeslice and the timeslice where the stochastic propagator is evaluated are identical. Therefore, the propagator is improved by removing terms in the corresponding hopping parameter expansion that are trivially zero in the exact case but contribute to the stochastic noise. A similar improvement is performed for the loop $L_1$ appearing in $S_1$ and $D$, which is also evaluated using stochastic propagators. Notice that for the $D$ contraction we have two versions, one using stochastic sources for both loops, and one where we use point sources only for one of the two loops. More details on the techniques are given in \cite{Bali:2021gel}. This reference gives also explicit expressions for the quantities that are evaluated on the lattice, except for the contraction $C_{1,duud}$, which contributes only in the context of flavor interference. We find:
\begin{align}
C_{1,duud}^{ij,\mvec{p}}(\mvec{y},t,\tau) &= 
	\frac{a^3}{ N_{\mathrm{st}}} 
	\left. 
		\sum_{\mvec{x}} \sum_{\ell}^{N_\mathrm{st}} 
		\left\langle \left[ 
			q_{2,t,j}^{T,\mvec{p},(\ell)}(x)\ 
			q_{1,t,i}^{\mvec{p},(\ell)}(x+y) 
		\right] \right\rangle 
	\right|_{x^4 = \tau, y^4 = 0}\,
\label{eq:def_c1}
\end{align}
with

\begin{align}
\label{eq:c1_q1_q2}
\left(q_{1,t,i}^{\mvec{p},(\ell)}\right)^a_{\alpha}(y) &:= 
	\left(\overline{S}_{321,t,i}^{\mvec{p},(\ell)}\right)^a_{\alpha}(y) -
	\left(\overline{S}_{123,t,i}^{\mvec{p},(\ell)}\right)^a_{\alpha}(y)
\,, \nonumber \\
\left(q_{2,t,j}^{\mvec{p},(\ell)}\right)^a_{\alpha}(y) &:= 
	\left[ 
		\psi_t^{\dagger,(\ell)}(y)\ 
		\gamma_5 \Gamma_j\ M^{\Phi ,\mvec{p}}_z(y) 
	\right]^{a}_{\alpha}\,,
\end{align}
where $\overline{S}$ is defined in equation (A.19) of \cite{Bali:2021gel}.

\paragraph{Lattice Setup:} We extend our simulation of reference \cite{Bali:2021gel} using the same lattice setup. In our simulation we employ the CLS ensemble H102 with $n_f = 2 + 1$ dynamical Sheikholeslami-Wohlert fermions \cite{Bruno:2016plf,Bruno:2014jqa}. For completeness we list again the corresponding lattice parameters in \tab\ref{tab:cls}. 

\begin{table}
\begin{center}
\begin{tabular}{ccccccccccc}
\hline
\hline
id & $\beta$ & $a[\mathrm{fm}]$  & $L^3 \times T$ & $\kappa_{l}$ & $\kappa_{s}$ & $m_{\pi}[\mathrm{MeV}]$ & $m_{K}[\mathrm{MeV}]$ & $m_\pi L a$ \\
\hline
H102 & $3.4$ & $0.0856$ & $32^3 \times 96$ & $0.136865$ & $0.136549339$ & $355$ & $441$ & $4.9$ \\
\hline
\hline
\end{tabular}
\end{center}
\caption{Details on the gauge ensemble H102, which is employed for our simulation \cite{Bruno:2016plf,Bruno:2014jqa}. We use 990 configurations.\label{tab:cls}}
\end{table}

In addition to the contractions $C_{1,uudd}$, $C_{1,uuuu}$, $C_{2,u}$, $C_{2,d}$, $S_{1,u}$, $S_{1,d}$, $S_{2}$ and $D$, which have been already calculated, we compute the contraction $C_{1,duud}$ according to \eqref{eq:def_c1} for proton momentum $\mvec{p} = \mvec{0}$. Like for the other $C_{1}$ contractions, we choose as source-sink separation $t = 12a$ and evaluate the four-point functions for insertion times $\tau \in [3a,t-3a]$. The corresponding data is fitted to a constant behavior in $\tau$.

\section{Results for invariant functions}
\label{sec:results}

\subsection{Data quality}

Before we discuss our results in a physical context, we first investigate the quality of our data. In particular, we focus on the potential presence of excited states and of lattice artifacts, such as anisotropy effects. We only treat the contraction $C_{1,duud}$, since all other contributions have been already investigated in our previous work \cite{Zimmermann:2019quf,Bali:2021gel}.

\begin{figure}
\begin{center}
\includegraphics[scale=0.35,trim={0.2cm 0.2cm 0.2cm 2.8cm},clip]{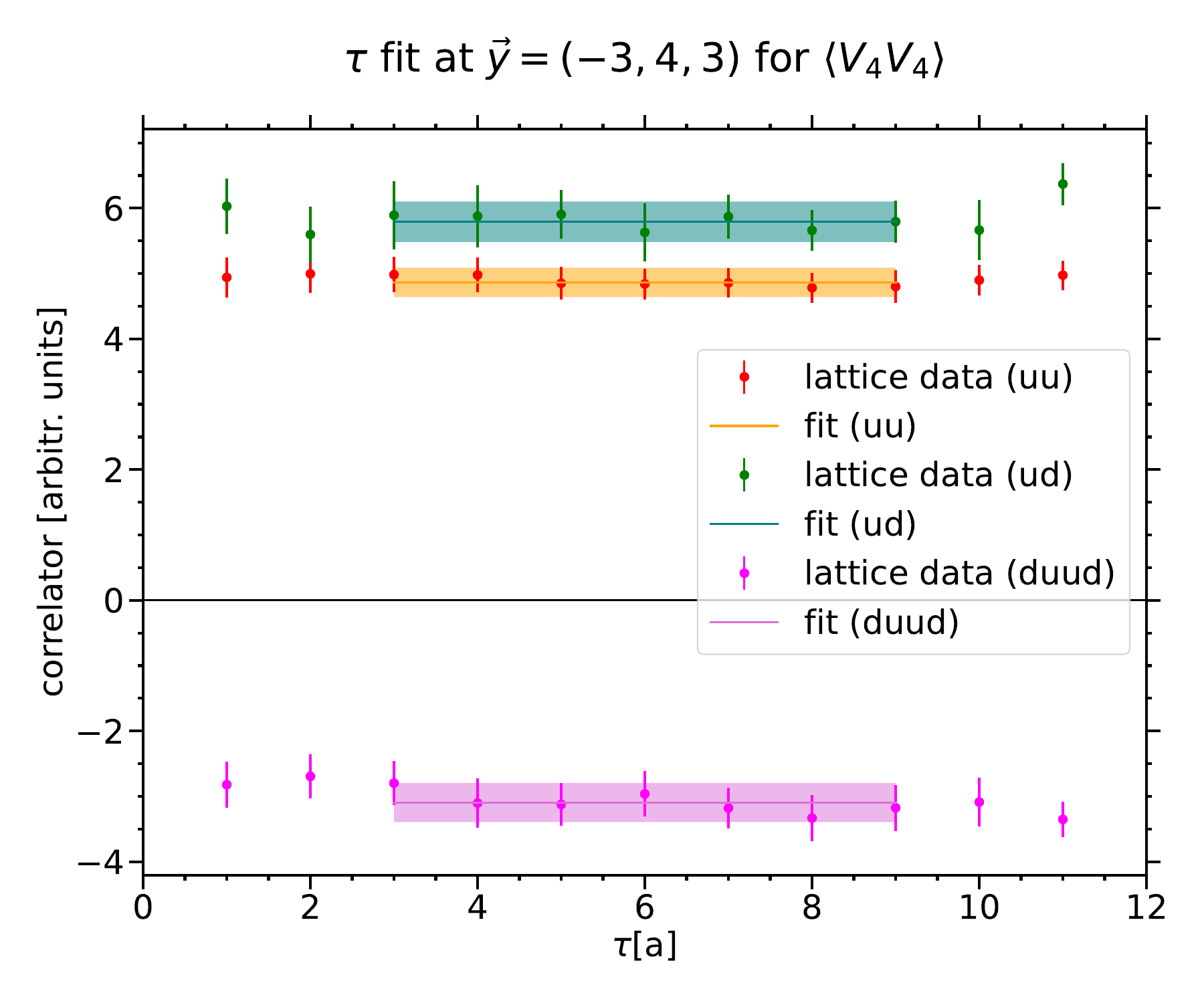}
\end{center}
\caption{Dependence on the insertion time $\tau$ of the contractions $C_{1,uuuu}$ (green), $C_{1,uudd}$ (red), and $C_{1,duud}$ (magenta) for the current combination $V_0 V_0$ and fixed quark separation $\mvec{y} = (-3,4,3)$. The orange band represents the constant fit that returns the final value for the given $\mvec{y}$. \label{fig:tau-platteau}}
\end{figure}

\paragraph{Excited states:} In the following we consider the $\tau$-dependence of the four-point correlator $C_{4\mathrm{pt}}(t,\tau,\mvec{y})$ defined in \eqref{eq:4ptdef} for fixed $\mvec{y}$. If there is a contamination of the data by excited states, this leads to a curvature along $\tau$. It turns out that there is in fact no visible dependence of the data on $\tau$. As an example we show the data of the $\langle V_0 V_0 \rangle$ correlator for the quark separation $\mvec{y} = (-3,4,3)a$ in \fig\ref{fig:tau-platteau}. The absence of curvature shows that excited states are sufficiently suppressed at the scale of our statistical uncertainties. The final value for the two-current correlator ground state is obtained by a constant fit in the region $3 \le \tau \le 9$. It has a reasonably small error and is also represented in \fig\ref{fig:tau-platteau} by the magenta band.

\begin{figure}
\begin{center}
\includegraphics[scale=0.35,trim={0.2cm 0.2cm 0.2cm 2.8cm},clip]{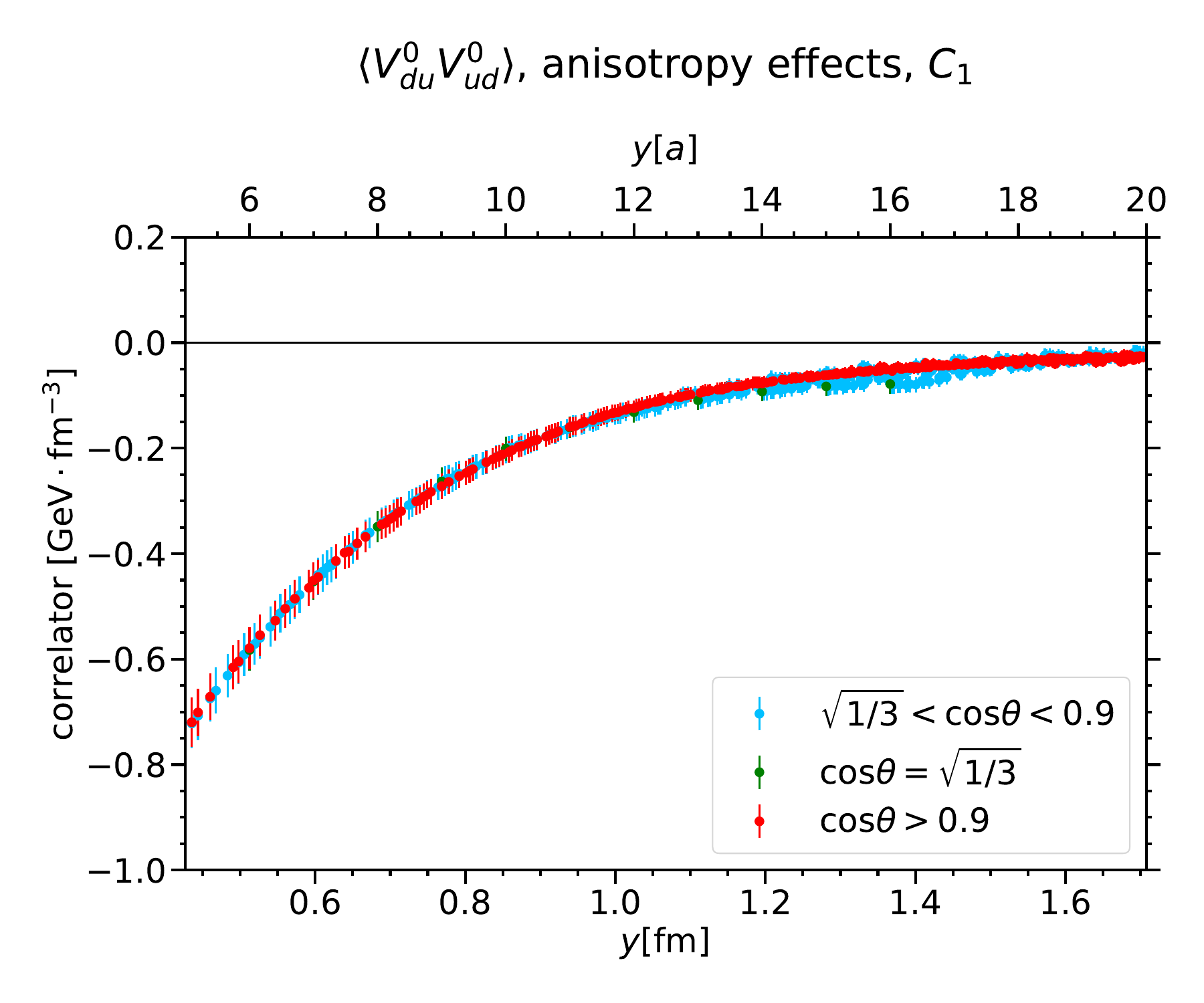}
\end{center}
\caption{The anisotropy of the $C_{1,duud}$ contribution to $\langle V_0 V_0 \rangle$, revealed by a saw tooth pattern for larger angles $\theta$ between the $\vec{y}$ vector and the nearest lattice diagonal.\label{fig:anisoduud}}
\end{figure}

\paragraph{Anisotropy:} Since we compute on a lattice with periodic boundary conditions, the effect of so-called ``mirror charges'' becomes important if the distance between the two current approaches half the lattice size \cite{Burkardt:1994pw,Alexandrou:2008ru}. As a consequence, there are anisotropy effects whose size depends on the angle $\theta(\vec{y}\ms)$ between $\vec{y}$ and the closest diagonal of the lattice. Moreover, the lattice propagator itself exhibits anisotropy effects, which become large at small distances \cite{Bali:2018spj,Cichy:2012is}. This is the case for contractions where both operators are connected directly via a quark propagator, i.e.\ the $C_2$ and $S_2$ contributions. In order to reduce the violation of Lorentz invariance, only regions with $|\vec{y}\ms|= y\ge 4a$ are considered. As discussed in \sect 4.1 of \cite{Bali:2021gel}, we limit ourselves to distances $y \le 16 a$ and take into account only data points close to the lattice diagonal, in order to reduce discretization artifacts and finite volumes effects. Specifically, only data points which fulfill 

\begin{align}
\cos(\theta(\vec{y}\ms))>0.9
\label{eq:diag-angle}
\end{align}
are considered. For the contraction $C_{1,duud}$ the most important source of anisotropy is given by mirror charges, which leads to the saw tooth pattern already observed in other contractions with $C_1$ topology. This is illustrated in \fig\ref{fig:anisoduud}, where the data points fulfilling the constraint \eqref{eq:diag-angle} are plotted in red.

\subsection{Invariant functions}

\begin{figure}
\begin{center}
\subfigure[$A_{qq^\prime}$ \label{fig:qq}]{
\includegraphics[scale=0.26,trim={0.2cm 0.2cm 0.2cm 2.8cm},clip]{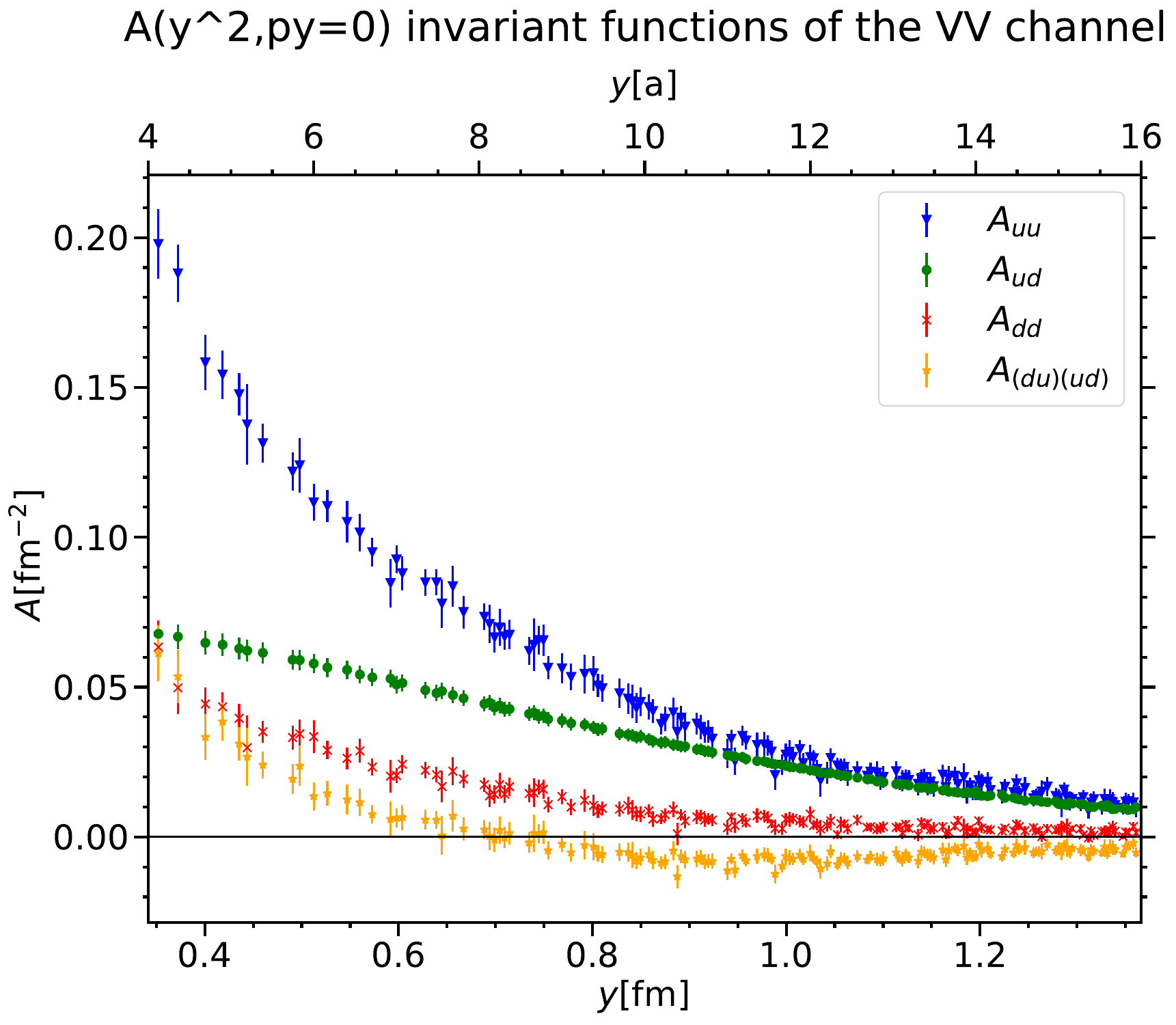}
}\hfill
\subfigure[$A_{q \delta q^\prime}$ \label{fig:qdq}]{
\includegraphics[scale=0.26,trim={0.2cm 0.2cm 0.2cm 2.8cm},clip]{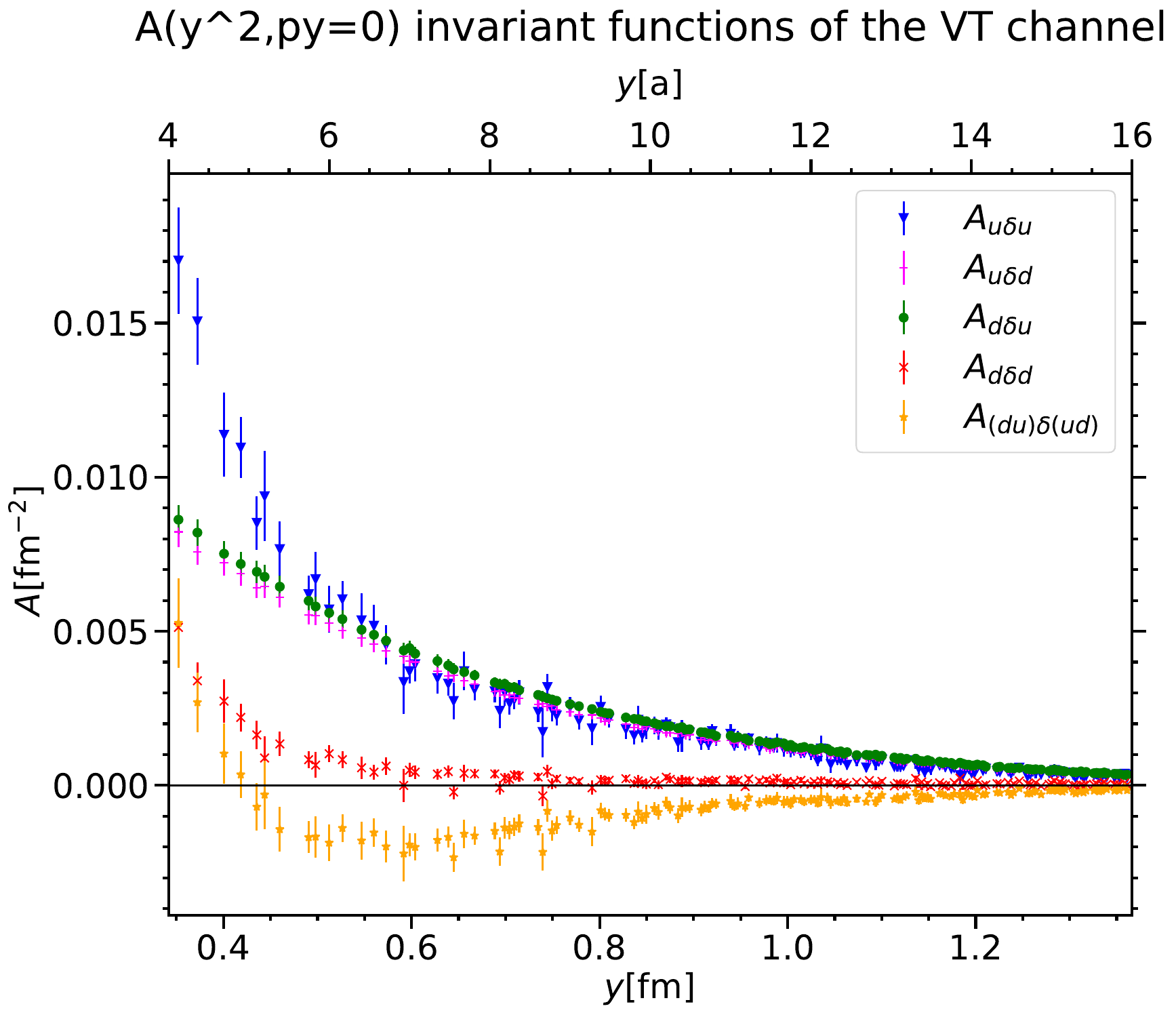}
}\\
\subfigure[$A_{\delta q \delta q^\prime}$ \label{fig:dqdq}]{
\includegraphics[scale=0.26,trim={0.2cm 0.2cm 0.2cm 2.8cm},clip]{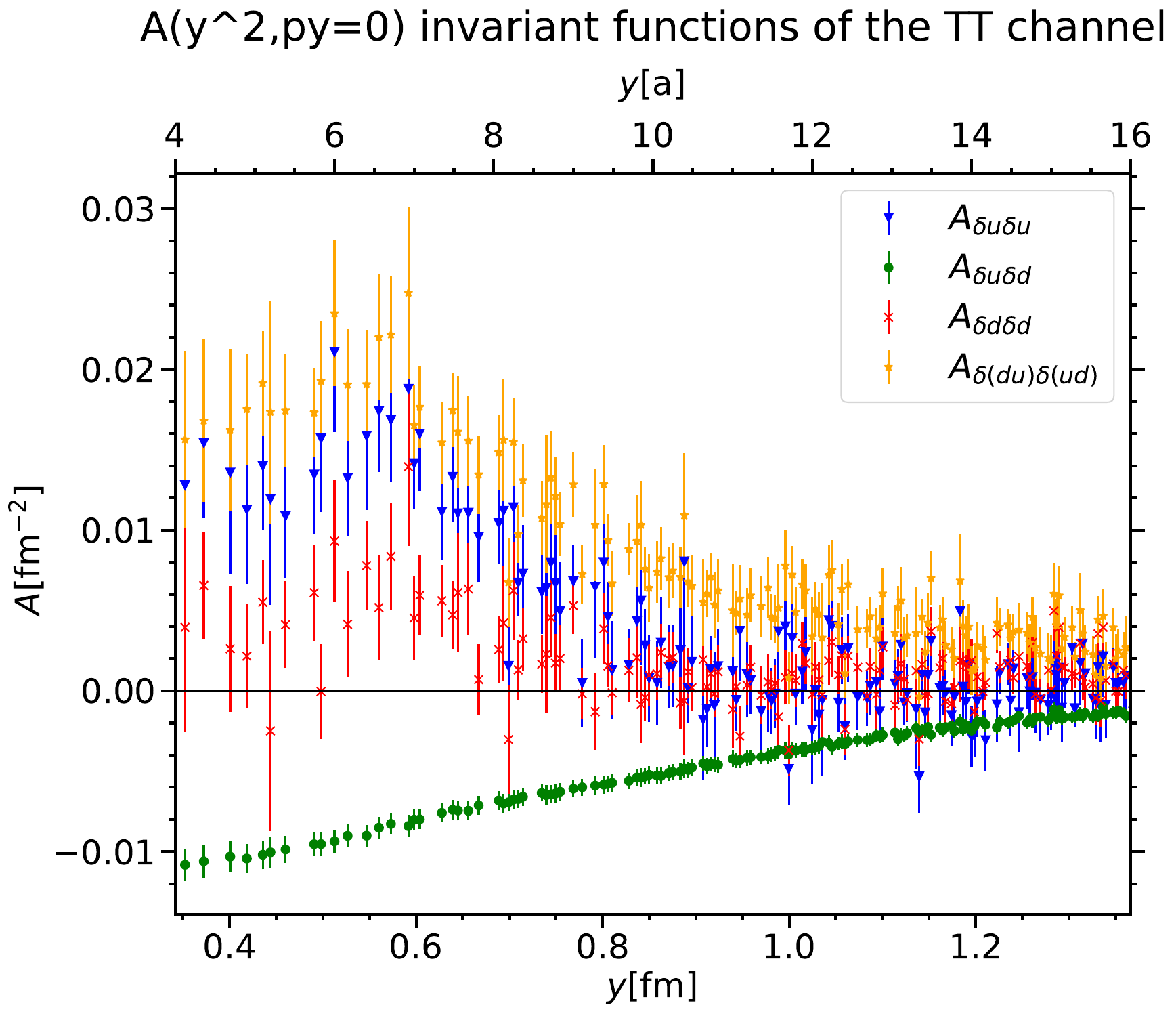}
}
\end{center}
\caption{Twist-two functions for three different polarization combinations and the flavor combinations $uu$, $ud$, $dd$ and $(du)(ud)$. Only the graphs $C_1$ and $C_2$ are taken into account. The hadron momentum used is always zero, $\mvec{p}=\mvec{0}$.\label{fig:twist-overview}}
\end{figure}

In the following, we consider physical matrix elements given by \eqref{eq:phys_me_decomp}. We take only into account the connected contributions $C_1$ and $C_2$ for quark separations fulfilling $4a\le y\le 16a$ and $\cos(\theta(y))>0.9$. The disconnected contribution $S_1$ was found to be small (but noisy) compared to $C_1$ and $C_2$ in \cite{Bali:2021gel}, whereas the errors on $D$ were too large to make any useful statement about its size. As pointed out in \cite{Bali:2021gel}, $S_2$ is seen to violate Lorentz invariance for $y<7a$, whereas it is orders of magnitudes smaller than the other contractions for larger quark distances. For that reason, we do not take into account this contraction in our physical results. The twist-two functions $A_{a_1 a_2}$ and $B_{\delta q \delta q^\prime}$ are obtained from the data by solving the overdetermined equation system \eqref{eq:tensor-decomp} for momentum $\mvec{p}=\mvec{0}$. The system of equations is solved by $\chi^2$-minimization (see \sect 3.2 of \cite{Bali:2020mij}). The corresponding results are shown in \fig\ref{fig:twist-overview} for the functions $A_{qq^\prime}$, $A_{q\delta q^\prime}$, and $A_{\delta q \delta q^\prime}$.\footnote{Our flavor generic notation $q$ and $q^\prime$ for indices includes the flavor interference cases $(ud)$ and $(du)$.} Here we show the results from \cite{Bali:2021gel} for the flavor diagonal combinations $uu$, $ud$, and $dd$, and compare them to our new results for the flavor interference channel. The latter are significantly different from zero and tend to be of similar size as the $dd$ contribution. Moreover, in the case of $A_{qq^\prime}$ and $A_{q\delta q^\prime}$ we observe a change in sign for the $(du)(ud)$ combination as a function of $y$. Hence, we can conclude that flavor interference can indeed be sizable for DPDs. We note that our results for $B_{\delta q \delta q'}$ have rather large statistical uncertainties for all flavor combinations other than $u d$ and are hence not shown here.

\begin{figure}
\begin{center}
\subfigure[Ratio $A_{(du)(ud)}/A_{ud}$ \label{fig:duud_uu}]{
\includegraphics[scale=0.248,trim={0.5cm 1.2cm 0.5cm 2.8cm},clip]{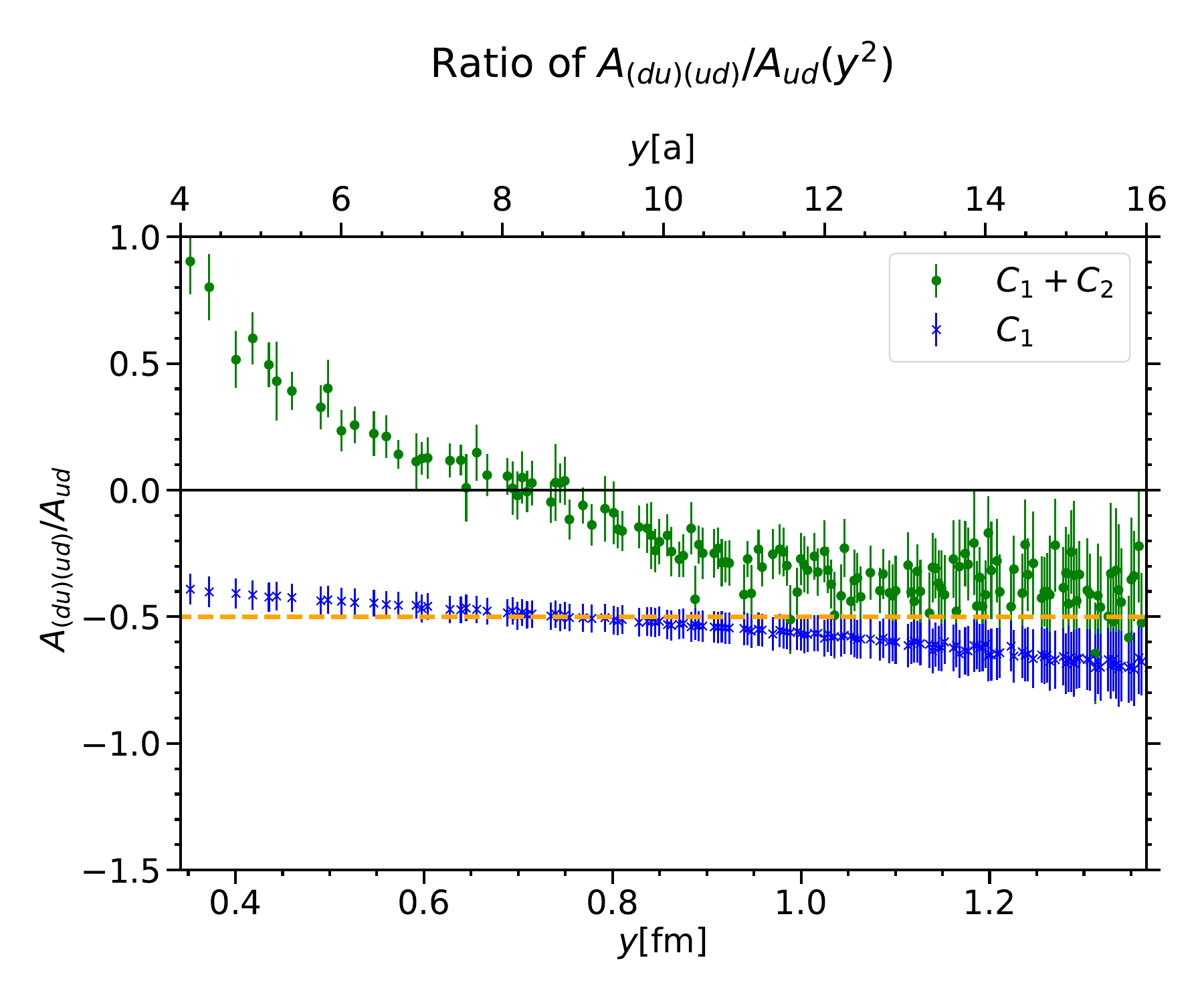}
}\hfill
\subfigure[Ratio $A_{(du)(ud)}/A_{uu}$ \label{fig:duud_ud}]{
\includegraphics[scale=0.248,trim={0.5cm 1.2cm 0.5cm 2.8cm},clip]{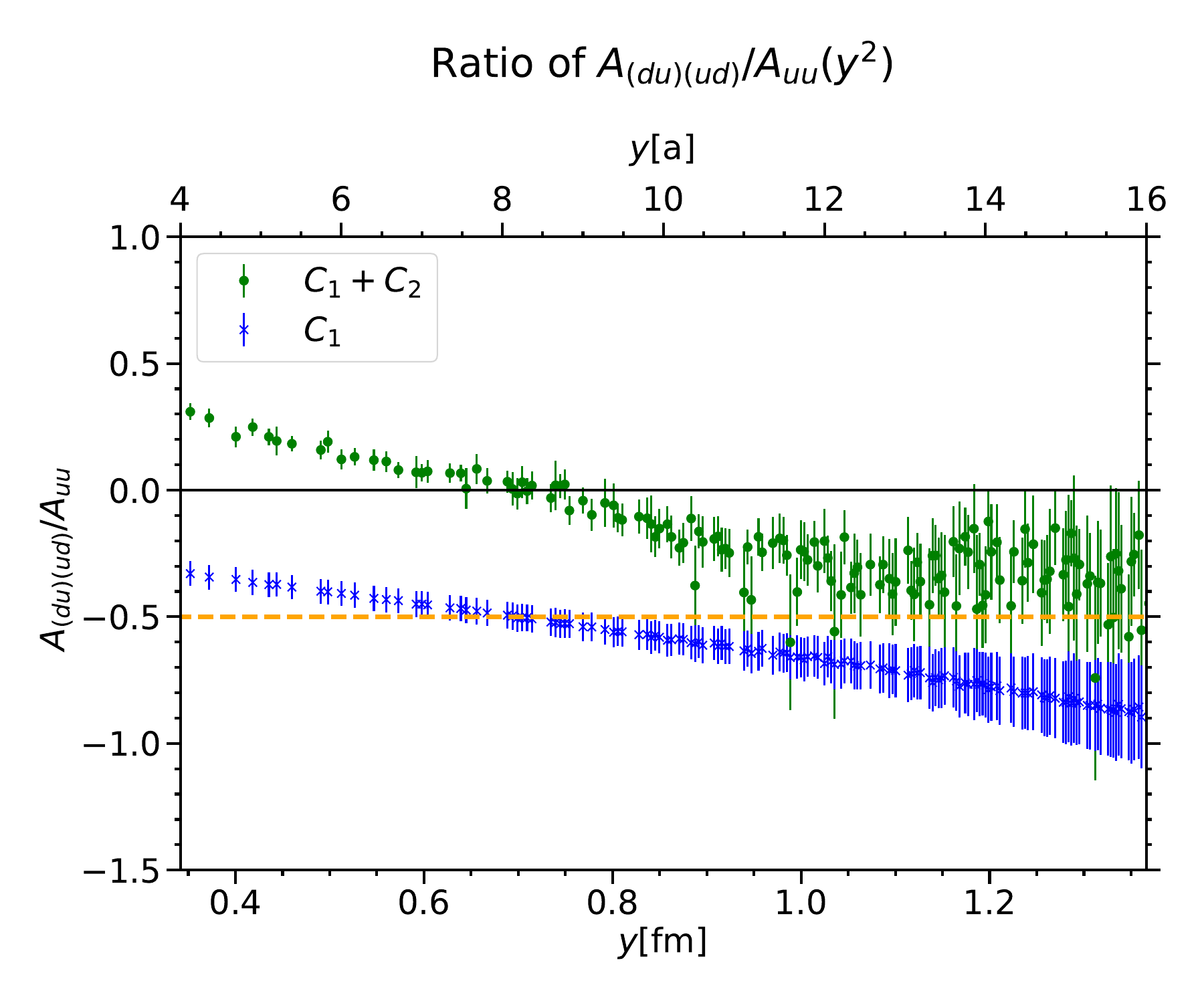}
}\\
\subfigure[Ratio $A_{uu}/A_{ud}$ \label{fig:uu_ud}]{
\includegraphics[scale=0.248,trim={0.5cm 1.2cm 0.5cm 2.8cm},clip]{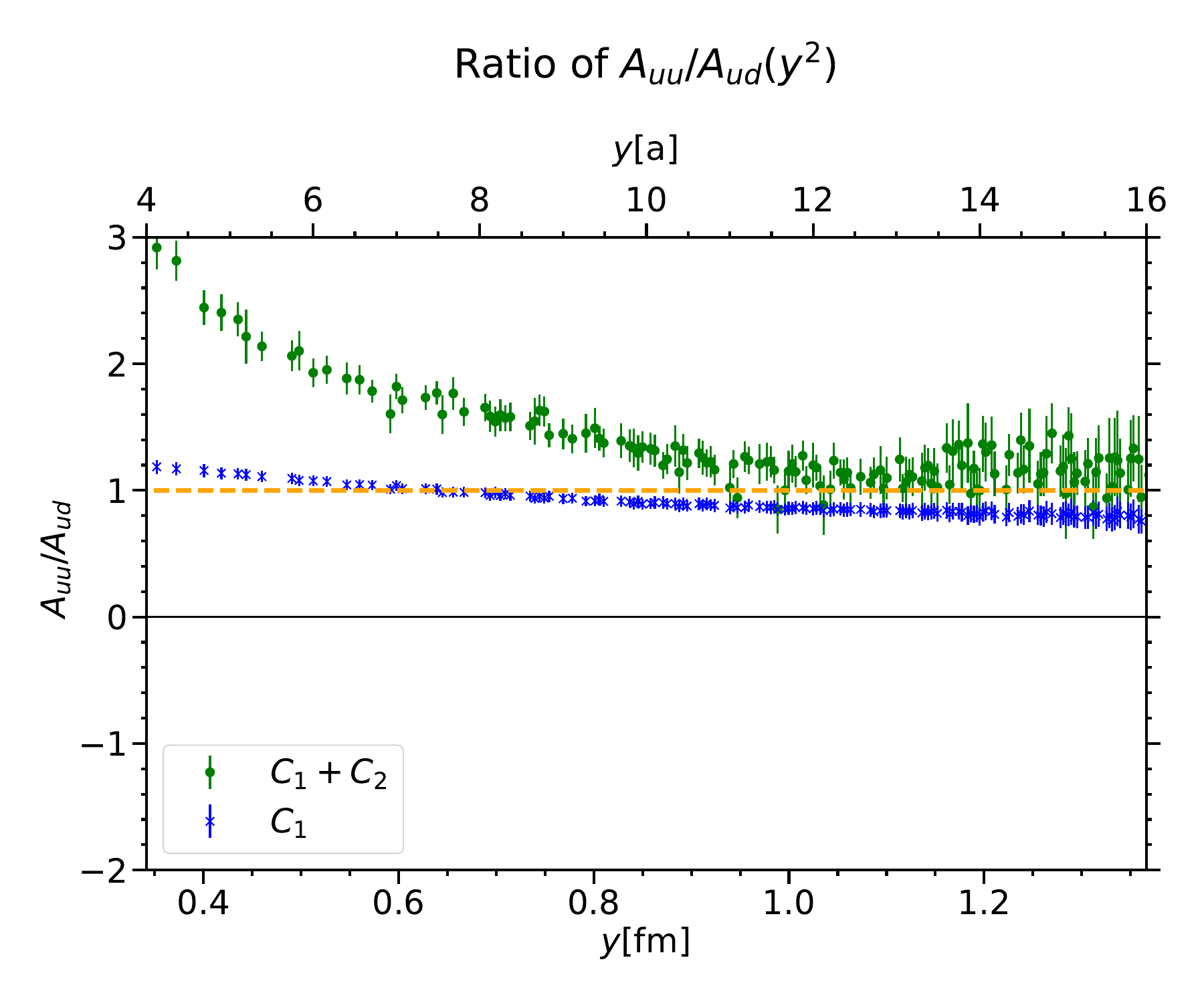}
}
\end{center}
\caption{Ratios of unpolarized twist-two functions. The blue points correspond to only considering the $C_1$ graphs, and the green ones to including both $C_1$ and $C_2$ contributions. The orange dashed line is the $SU(6)$ model prediction.\label{fig:ratio-overview}}
\end{figure}

\begin{figure}
\begin{center}
\subfigure[Ratio $A_{\delta(du)\delta(ud)}/A_{ud}$ \label{fig:TTduud_ud}]{
\includegraphics[scale=0.248,trim={0.5cm 1.2cm 0.5cm 2.8cm},clip]{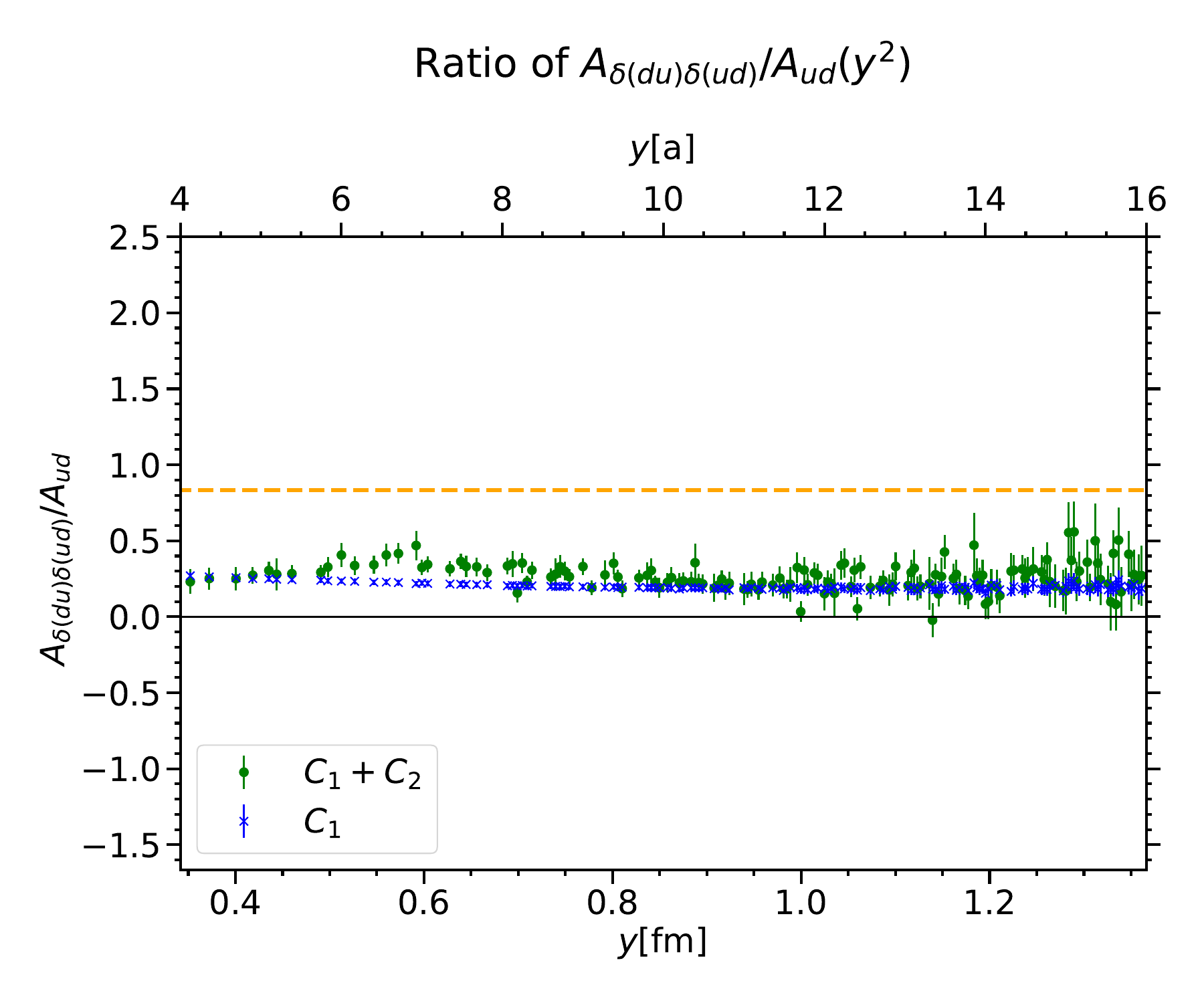}
}\hfill
\subfigure[Ratio $A_{\Delta(du)\Delta(ud)}/A_{ud}$ \label{fig:AAduud_ud}]{
\includegraphics[scale=0.248,trim={0.5cm 1.2cm 0.5cm 2.8cm},clip]{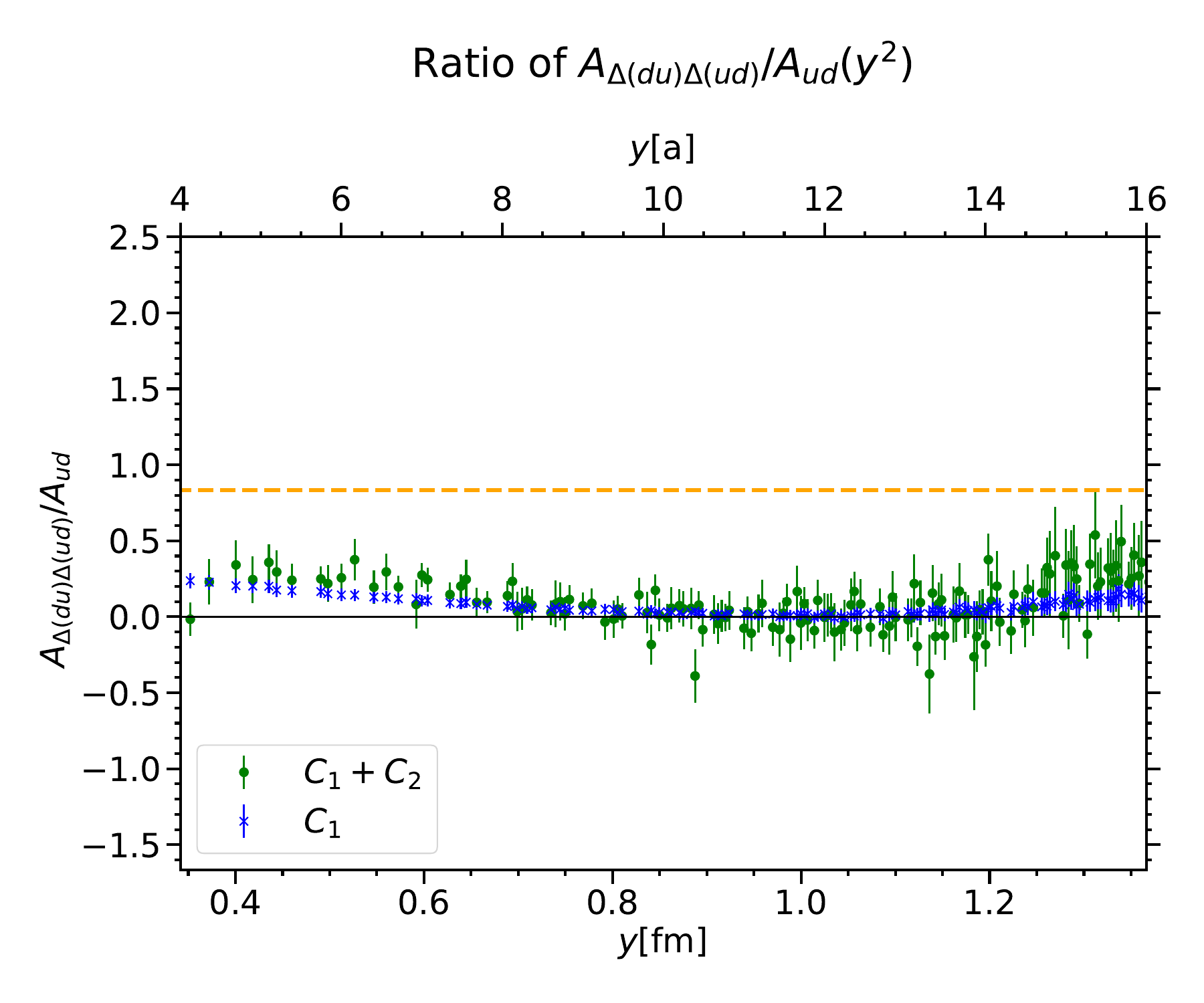}
} \\
\subfigure[Ratio $A_{\Delta u\Delta d}/A_{ud}$ \label{fig:AAud_ud}]{
\includegraphics[scale=0.248,trim={0.5cm 1.2cm 0.5cm 2.8cm},clip]{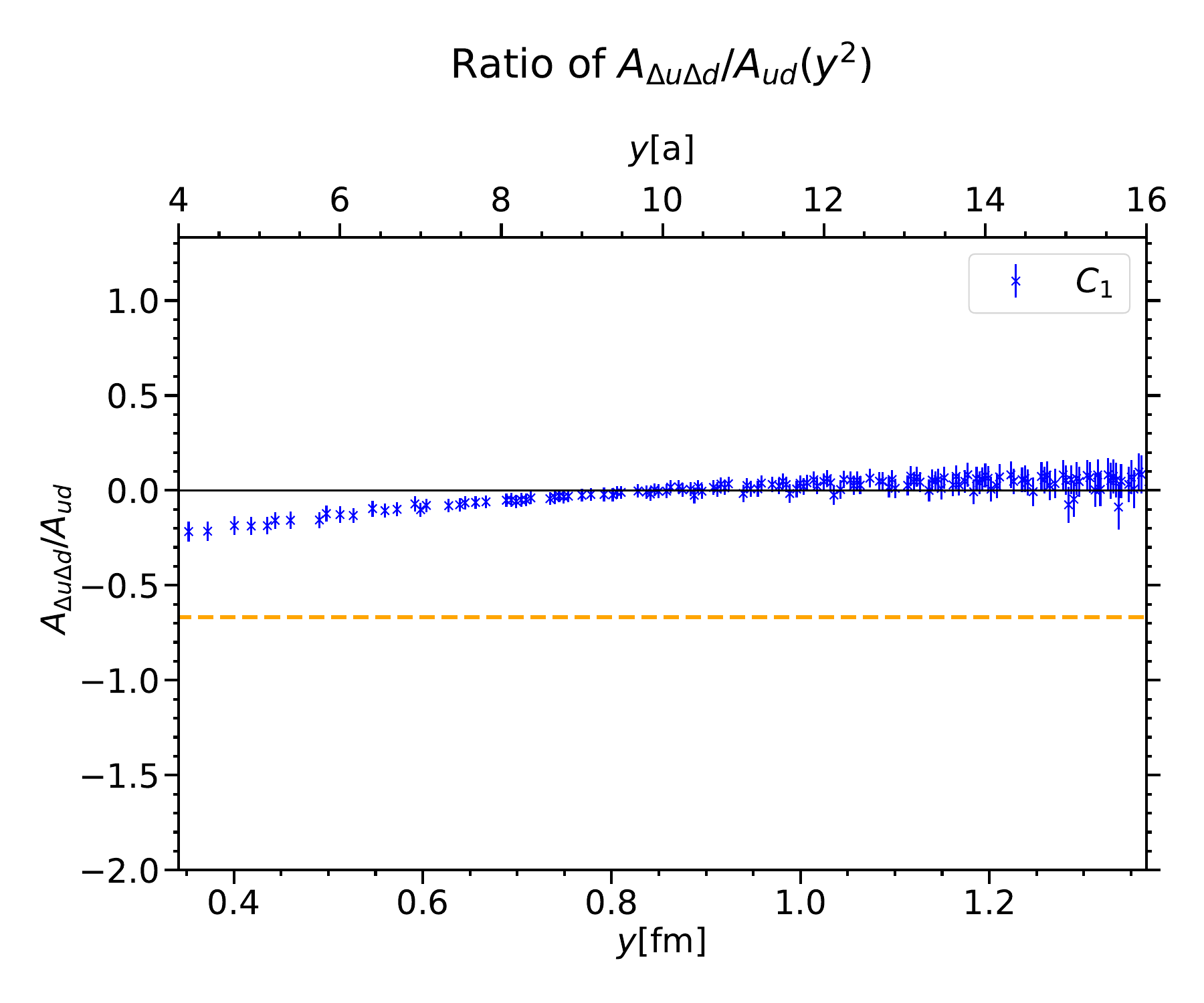}
}\hfill
\subfigure[Ratio $A_{\delta u\delta d}/A_{ud}$ \label{fig:TTud_ud}]{
\includegraphics[scale=0.248,trim={0.5cm 1.2cm 0.5cm 2.8cm},clip]{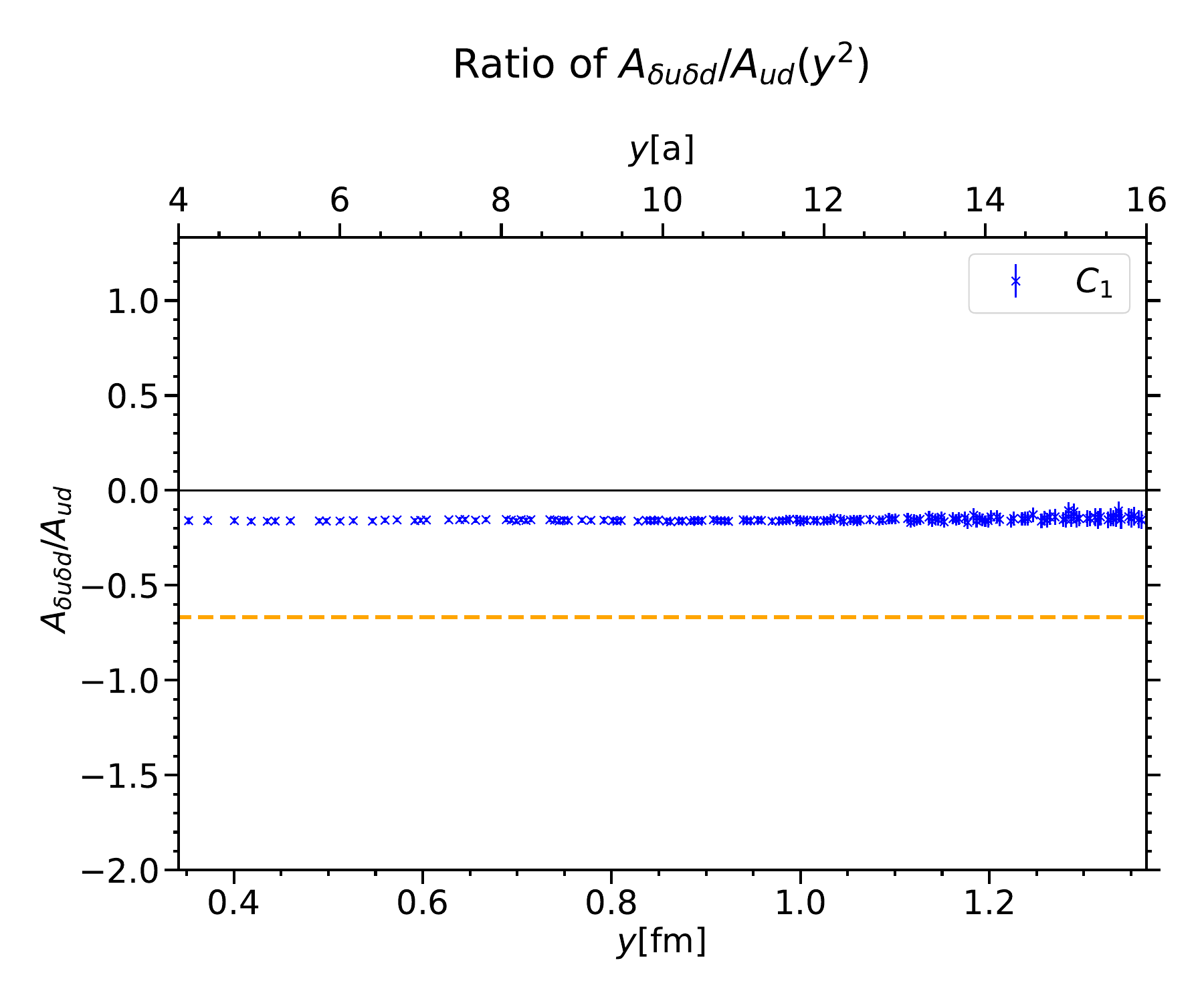}
}\\
\subfigure[Ratio $A_{\Delta u\Delta u}/A_{uu}$ \label{fig:AAuu_uu}]{
\includegraphics[scale=0.248,trim={0.5cm 1.2cm 0.5cm 2.8cm},clip]{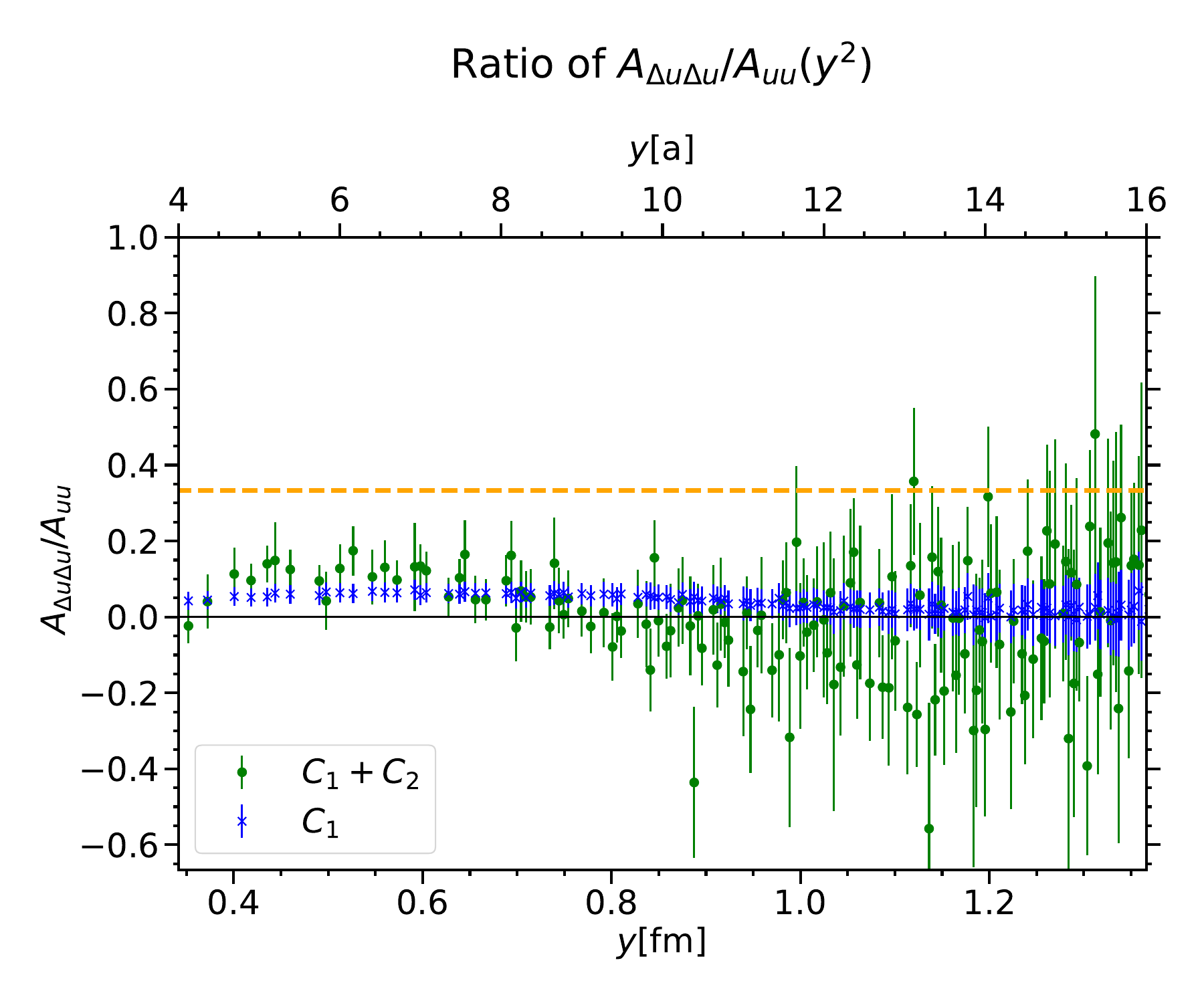}
}\hfill
\subfigure[Ratio $A_{\delta u\delta u}/A_{uu}$ \label{fig:TTuu_uu}]{
\includegraphics[scale=0.248,trim={0.5cm 1.2cm 0.5cm 2.8cm},clip]{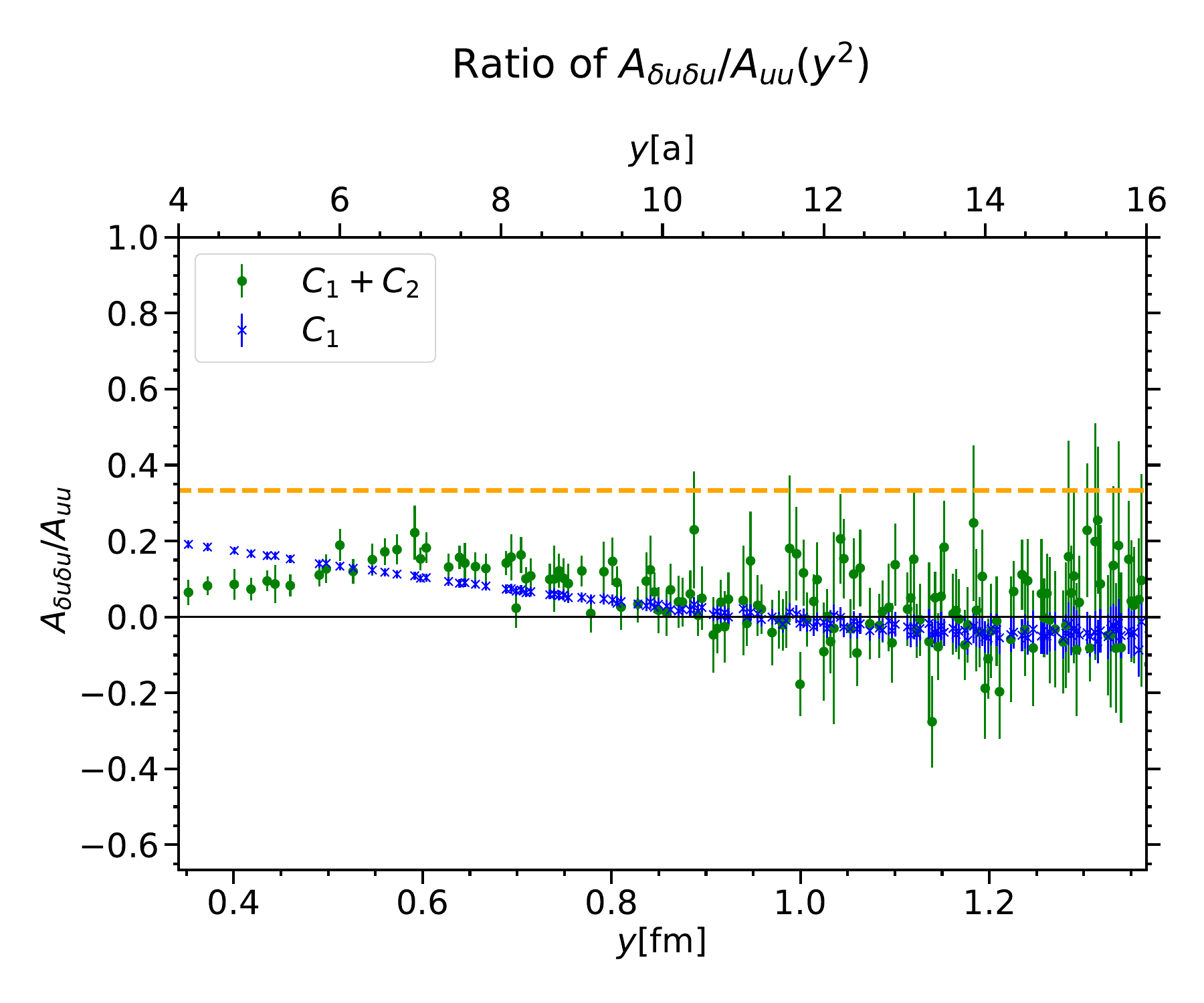}
}
\end{center}
\caption{As \fig\ref{fig:ratio-overview}, but for ratios of polarized and unpolarized functions. Note that there is no $C_2$
contribution to the flavor combination $u d$ in panels (c) and
(d)\label{fig:ratio-pol-overview}}
\end{figure}

\subsection{Comparison with $SU(6)$ predictions}

In the following, we compare the ratios \eqref{eq:su6-ratios} derived for an $SU(6)$-symmetric three-quark wave function with the corresponding data we obtain from the lattice. Note that the evaluation of the DPDs from a three-quark wave function corresponds to the contraction $C_1$. whereas the interpretation of $C_2$ involves wave functions with additional quark-antiquark pairs, as discussed in \sect 4.1 of \cite{Bali:2020mij}. We therefore compare the $SU(6)$ predictions with either $C_1$ alone, or with the sum of $C_1$ and $C_2$. This is shown in \fig\ref{fig:ratio-overview} for the unpolarized flavor ratios \eqref{eq:su6-ratios} and in \fig\ref{fig:ratio-pol-overview} for the polarization ratios \eqref{eq:su6-ratios-pol}. 

In the case of two unpolarized quarks, the $SU(6)$ model results are fairly consistent with the corresponding $C_1$ results. For large quark distances $y$, this also holds for the sum of $C_1$ and $C_2$. However, for smaller $y$, discrepancies become rather large. By contrast, lattice results for polarization ratios, which are shown in \fig\ref{fig:ratio-pol-overview}, disagree strongly with the $SU(6)$ model.

\section{Factorization tests}
\label{sec:factorization}

A common strategy for modeling DPDs is to express them in terms of single-parton distributions, assuming that correlations between the two partons can be neglected. A corresponding ansatz can also be formulated at the level of twist-two functions \cite{Bali:2020mij,Bali:2021gel}. We now extend this formulation to the flavor interference case, which we will find to be special in this context. We limit ourselves to unpolarized quarks in the following

Technically, the factorization ansatz is obtained by inserting a complete set of eigenstates between the two operators in the DPD matrix element in \eqref{eq:dpd-def} and neglecting all intermediate states except for the ground state:

\begin{align}
&\avsum_{\lambda} \bra{p,\lambda} \Op_{a_1}(y,z_1)\ \Op_{a_2}(0,z_2) \ket{p,\lambda} \nonumber \\
&\quad\approx \avsum_{\lambda,\lambda^\prime} 
\int 
\frac{\dd p^{\prime +}\, \dd^2 \tvec{p}^\prime}{2 p^{\prime +} (2\pi)^3} \,
e^{-iy(p^\prime - p)} 
\bra{p,\lambda} \Op_{a_1}(0,z_1) \ket{p^\prime,\lambda^\prime} 
\bra{p^\prime,\lambda^\prime} \Op_{a_2}(0,z_2) \ket{p,\lambda}\,.
\label{eq:ans_f1}
\end{align}
In the flavor diagonal case, the matrix elements on the r.h.s.\ can be directly identified with GPD matrix elements $f_a^{\lambda\lambda^\prime}(\bar{x},\xi,\tvec{p}^\prime,\tvec{p})$, which leads to the factorization formula for DPDs \cite{Diehl:2011yj}:

\begin{align}
&F_{a_1 a_2}(x_1,x_2,\zeta,\tvec{y})
\stackrel{?}{=}
\frac{1}{2(1-\zeta)}
\int \frac{\dd^2 \tvec{r}}{(2\pi)^2} \,
e^{-i\tvec{r}\tvec{y}}
\sum_{\lambda \lambda^\prime}
f_{a_1}^{\lambda\lambda^\prime}(\bar{x}_1,-\xi,\tvec{0},-\tvec{r})\
f_{a_2}^{\lambda^\prime \lambda}(\bar{x}_2,\xi,-\tvec{r},\tvec{0})\,,
\label{eq:dpd-fact-diag}
\end{align}
where $f$ is defined in equation (5.5) of \cite{Bali:2021gel}. The notation $\stackrel{?}{=}$ indicates that \eqref{eq:dpd-fact-diag} is an assumption, which we investigate in the following. $\bar{x}_i$ and $\xi$ are functions of $\zeta$ and/or $x_i$:

\begin{align}
\bar{x}_i(x_i,\zeta) := \frac{2x_i}{2-\zeta} \,, \qquad  \xi(\zeta) := \frac{\zeta}{2-\zeta} \,.
\end{align}
\begin{figure}
\begin{center}
\subfigure[$F_{ud}$]{
\includegraphics[scale=0.47,trim={0.5cm 0cm 0.5cm 0cm},clip]{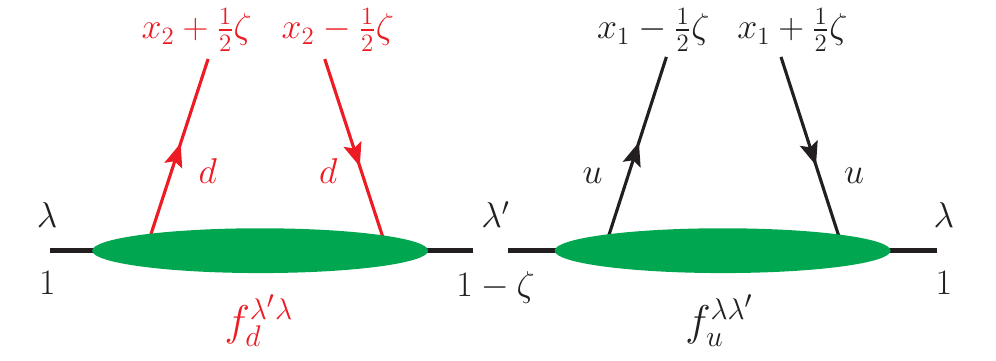}
}\hfill
\subfigure[$F_{(ud)(du)}$]{
\includegraphics[scale=0.47,trim={0.5cm 0cm 0.5cm 0cm},clip]{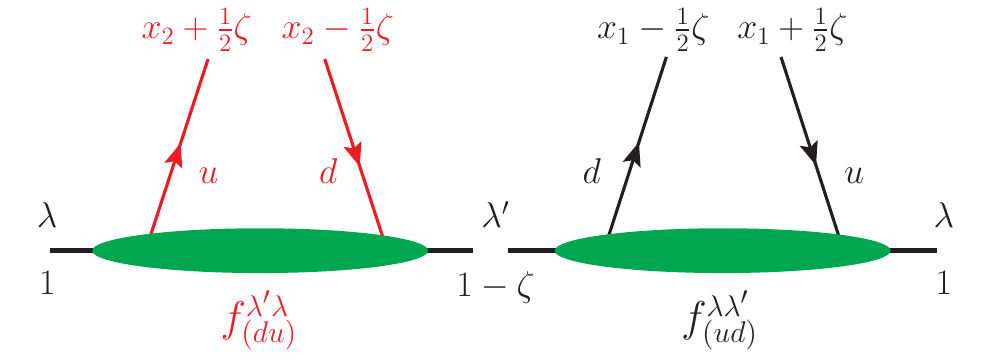}
}
\end{center}
\caption{Depiction of the factorized expression \eqref{eq:dpd-fact-diag} for the flavor-diagonal case (a), as well as the interference case (b).\label{fig:fact-graph}}
\end{figure}
\begin{figure}
\begin{center}
\includegraphics[scale=1]{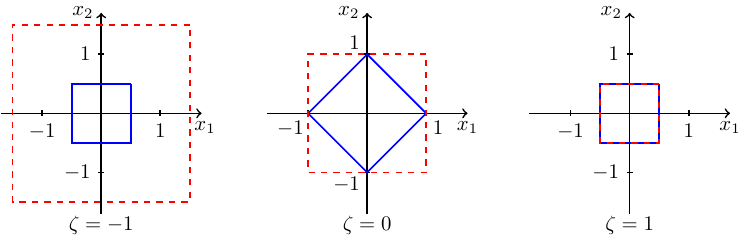}
\end{center}
\caption{Support regions for different values of $\zeta$, blue: physical support region of the DPD, red dashed: support region of the factorized expression\label{fig:support}}
\end{figure} 
For flavor-changing operators, the matrix elements in the second line of \eqref{eq:ans_f1} are related to GPDs for the transition between a proton and a neutron. In \fig\ref{fig:fact-graph} we show a pictorial representation of the factorized expression in the flavor diagonal and non-diagonal cases. Using isospin symmetry, one can relate these transition GPDs to ordinary ones \cite{Mankiewicz:1997aa}: 

\begin{align}
\bra{p} \Op_{(ud)} \ket{n} 
= 
\bra{p} \Op_u \ket{p} - \bra{p} \Op_d \ket{p}
=
\bra{n} \Op_{(du)} \ket{p} \,,
\end{align}
which leads to the following decomposition of the flavor interference matrix element in terms of flavor conserving matrix elements:

\begin{align}
&\bra{p} \Op_{(ud)} \ket{n} \bra{n} \Op_{(du)} \ket{p}\ 
=
\Bigl[\bra{p} \Op_{u} \ket{p} - \bra{p} \Op_{d}  \ket{p}\Bigr]^2
\nonumber \\
&\quad =
\bra{p} \Op_{u} \ket{p} \bra{p} \Op_{u} \ket{p} 
- 2	\bra{p} 	\Op_{u} \ket{p} \bra{p} \Op_{d} \ket{p}
+ \bra{p} \Op_{d} \ket{p} \bra{p} \Op_{d} \ket{p} \,. 
\label{eq:trans-me-iso-decomp}
\end{align}
Inserting this into the factorization hypothesis \eqref{eq:ans_f1}, we find that we can relate the flavor interference DPDs to a combination of flavor diagonal ones:

\begin{align}
F_{(ud)(du)}(x_1,x_2,\zeta,\tvec{y}) 
= F_{uu}(x_1,x_2,\zeta,\tvec{y}) - 2F_{ud}(x_1,x_2,\zeta,\tvec{y}) + F_{dd}(x_1,x_2,\zeta,\tvec{y}) \,.
\label{eq:F_fact}
\end{align}
Moreover, taking Mellin moments and performing a Fourier transform w.r.t.\ $\zeta$, we find for the invariant functions:

\begin{align}
A_{(ud)(du)}(py,y^2) = A_{uu}(py,y^2) - 2A_{ud}(py,y^2) + A_{dd}(py,y^2) \,.
\label{eq:A_iso}
\end{align}
Notice that the ordering of the operators is important. If we considered $\Op_{(du)}\, \Op_{(ud)}$ instead of $\Op_{(ud)}\, \Op_{(du)}$, the intermediate state with lowest energy would not be a nucleon and the corresponding transition GPDs could not be related with those in the proton. Hence, we are restricted to the operator ordering given in \eqref{eq:trans-me-iso-decomp}. In \cite{Bali:2021gel}, we pointed out that the regions of support of the two sides of \eqref{eq:dpd-fact-diag} differ to a degree that depends on the value of $\zeta$. In particular, the mismatch is worst for $\zeta < 0$ (see \fig\ref{fig:support}). Hence, we used \eqref{eq:dpd-fact-diag} only for $\zeta \ge 0$ whilst for $\zeta < 0$ we derived a factorized expression for the order $\Op_{a_2} \Op_{a_1}$. As we just explained, we cannot do that in the flavor interference case, where we have to keep one ordering of the operators and integrate over the entire $\zeta$-region. This leads to:

\begin{align}
&A_{(ud)(du)}(py=0,y^2) 
\stackrel{?}{=} 
\frac{1}{4\pi^2} \int_{-1}^1 \dd \zeta\; \frac{(1-\frac{\zeta}{2})^2}{2(1-\zeta)} 
\int \dd r\ r\, J_0 (y r) \sum_{\lambda \lambda^\prime} 
\int \dd x_1\ \int \dd x_2\ \nonumber \\
&\qquad\times 
\left[ 
f_{u}^{\lambda \lambda^\prime} ( x_1, -\xi, \tvec{0}, -\tvec{r} )  
f_{u}^{\lambda^\prime \lambda} ( x_2, \xi, -\tvec{r}, \tvec{0} ) -
2 f_{u}^{\lambda \lambda^\prime} ( x_1, -\xi, \tvec{0}, -\tvec{r} )  
f_{d}^{\lambda^\prime \lambda} ( x_2, \xi, -\tvec{r}, \tvec{0} ) 
\right. \nonumber \\ 
&\qquad\quad\left.+ 
f_{d}^{\lambda \lambda^\prime} ( x_1, -\xi, \tvec{0}, -\tvec{r} ) 
f_{d}^{\lambda^\prime \lambda} ( x_2, \xi, -\tvec{r}, \tvec{0} ) 
\right] \,,
\label{eq:fact_test_A}
\end{align}
where $y=|\tvec{y}|$ and $r=|\tvec{r}|$. Notice that the expressions in the square brackets are rotationally invariant w.r.t.\ to the momentum $\tvec{r}$, which allowed us to perform the angular part of the integration over $\tvec{r}$. After integration over $x_1$ and $x_2$, they can be expressed in terms of Pauli and Dirac form factors, $F_1$ and $F_2$. For the second term in the square brackets of \eqref{eq:fact_test_A}, we obtain:

\begin{align}
&\frac{1}{2} \sum_{\lambda\lambda^\prime} 
\int \dd x_1 \int \dd x_2\ 
f_{u}^{\lambda \lambda^\prime} ( x_1, -\xi, \tvec{0}, -\tvec{r} )\ 
f_{d}^{\lambda^\prime \lambda} ( x_2, \xi, -\tvec{r}, \tvec{0} ) 
\nonumber\\ 
&\qquad = 
K_1(\zeta) F_{1}^{u}(t) F_{1}^{d}(t) 
- 
K_2(\zeta) \left[ F_{1}^{u}(t) F_{2}^{d}(t) +  F_{1}^{d}(t) F_{2}^{u}(t) \right]
\nonumber\\
&\qquad\quad + 
\left( K_3(\zeta) + \frac{\tvec{r}^2}{4m^2} K_4(\zeta) \right) 
F_{2}^{u}(t) F_{2}^{d}(t) \,,
\end{align}
with

\begin{align}
t(\zeta,\tvec{r}^2) &:= -\frac{\zeta^2 m^2 + \tvec{r}^2}{1-\zeta} \,,
\qquad
K_1(\zeta) := 1-K_2(\zeta) \,,
\qquad 
K_2(\zeta) := \frac{\zeta^2}{(2-\zeta)^2} \,,
\nonumber \\
K_3(\zeta) &:= \frac{\bigl(K_2(\zeta)\bigr)^2}{K_1(\zeta)}\,,
\qquad 
K_4(\zeta) := \frac{1}{1-\zeta}\,.
\end{align}
Analogous expressions hold for the other terms.

For the Dirac and Pauli form factors, we take the results of lattice simulations obtained with the same ensemble used in the present work. We fitted these data to several parametrizations in \cite{Bali:2021gel}. In the following plots, bands correspond to the range obtained when inserting these different parametrizations into the factorization formula.

\begin{figure}[h]
\subfigure[$C_1+C_2$ for flavors $uu$]{
\includegraphics[scale=0.26,trim={0.2cm 0.2cm 0.2cm 2.8cm},clip]{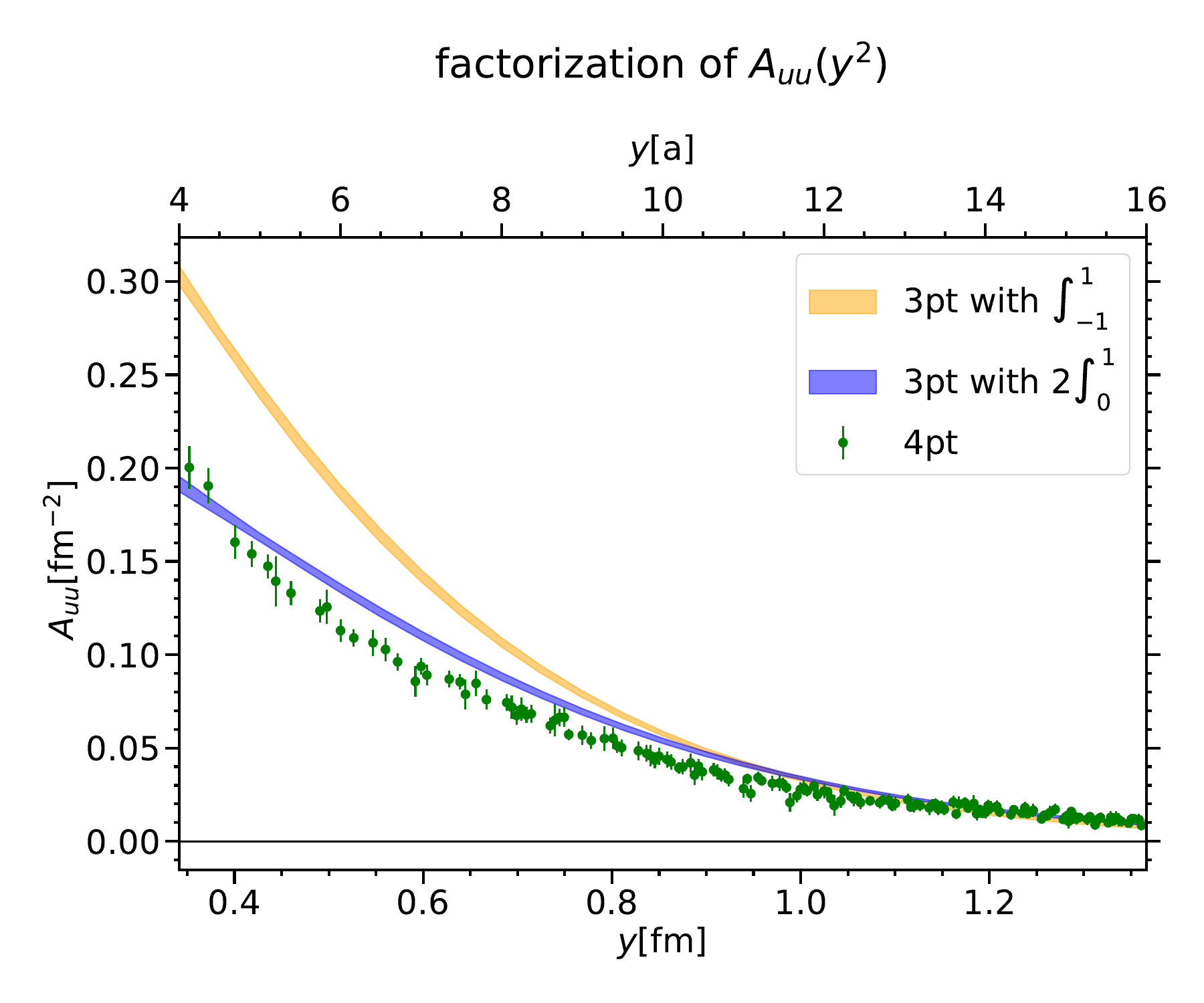}
}\hfill
\subfigure[$C_1+C_2$ for flavors $ud$]{
\includegraphics[scale=0.26,trim={0.2cm 0.2cm 0.2cm 2.8cm},clip]{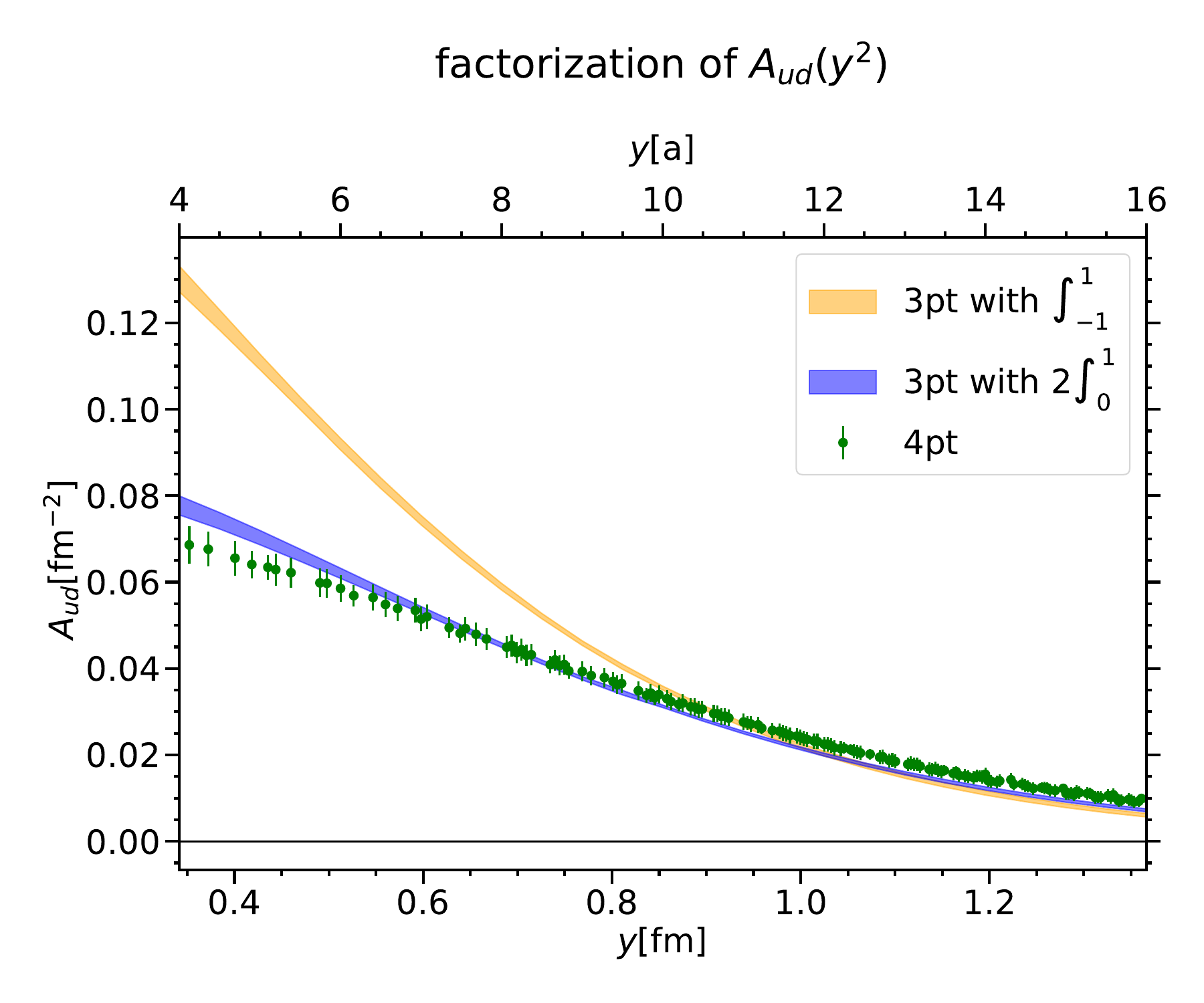}
}
\caption{Factorization of the unpolarized flavor conserving twist-two functions $A_{ud}$ and $A_{uu}$. The blue band shows the result for the factorization for the reduced support region that has been published in \cite{Bali:2021gel}. The orange band represents the factorization result where \eqref{eq:dpd-fact-diag} has been directly applied without exchanging the order of operators.\label{fig:fact_compar}}
\end{figure}

In order to check the impact of integrating \eqref{eq:ans_f1} over the entire $\zeta$-region instead of exchanging the order of operators, we repeat the factorization analysis of the flavor diagonal DPDs given in \cite{Bali:2021gel} by using the correspondingly modified ansatz. The result of this is shown in \fig\ref{fig:fact_compar}, along with our original result from \cite{Bali:2021gel}, where \eqref{eq:ans_f1} was only used for $\zeta \ge 0$. We see that the difference due to the increased unphysical support region is considerable when going to quark distances below about $0.8~\mathrm{fm}$. This shows that integrating \eqref{eq:ans_f1} over the entire $\zeta$-region yields a very poor approximation for small quark distances. Unfortunately, for the flavor interference contribution this is the only possibility we have found.

\begin{figure}[h]
\begin{center}
\includegraphics[scale=0.26,trim={0.2cm 0.2cm 0.2cm 2.8cm},clip]{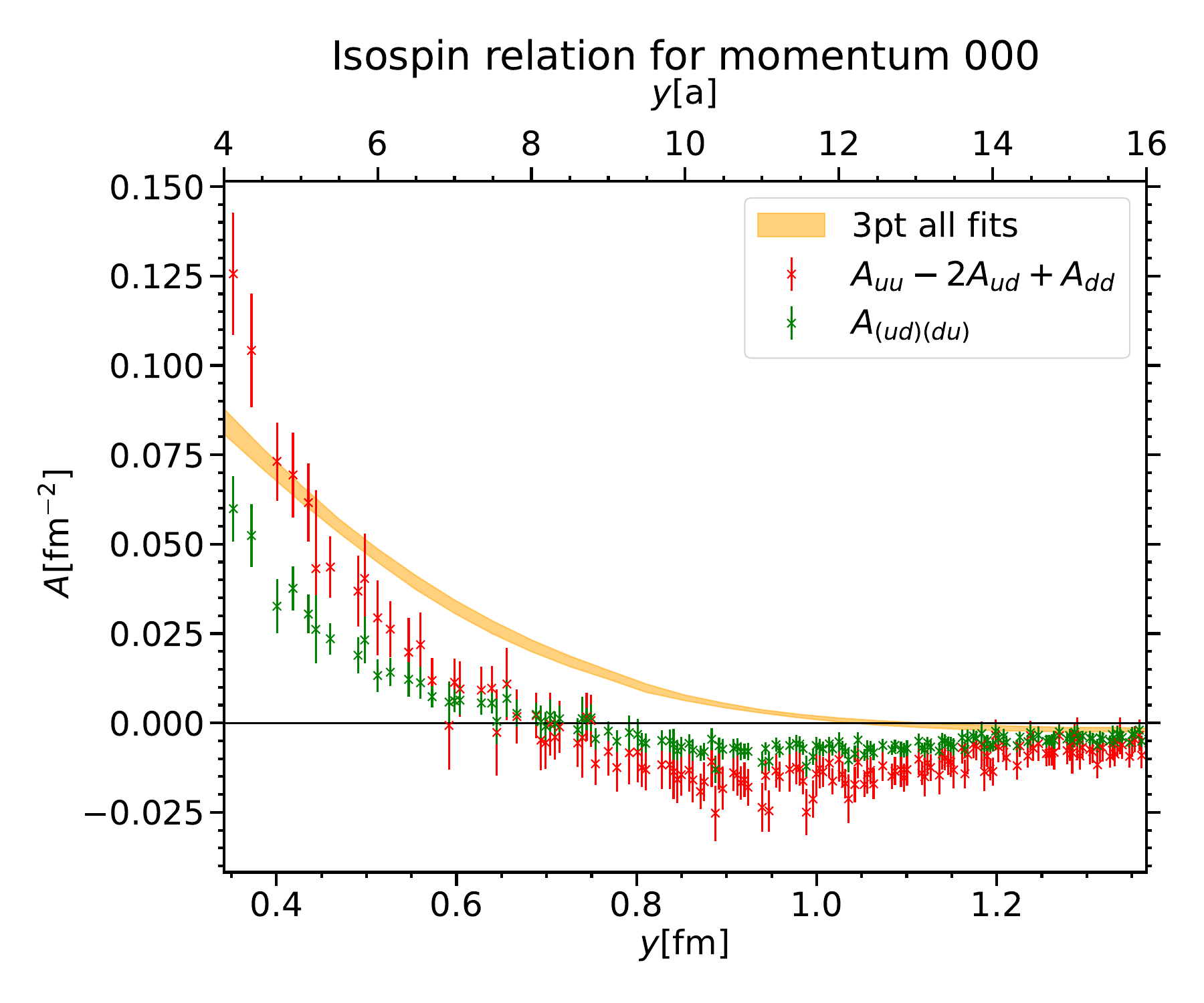}
\end{center}
\caption{Factorization of the unpolarized flavor interference twist-two function $A_{(ud)(du)}$. The orange band represents the result obtained from \eqref{eq:fact_test_A}. This is compared to the corresponding results obtained from our four-point function calculation. Here we show the data points for $A_{(ud)(du)}$ (green) and the r.h.s.\ of the flavor relation \eqref{eq:A_iso} (red).\label{fig:uddufact}}
\end{figure}

The result for flavor interference DPDs is plotted in \fig\ref{fig:uddufact} and compared to the corresponding result of the direct calculation (green). Based on the comparison in \fig\ref{fig:uddufact}, we consider the factorization formula that uses the full $\zeta$-region as unreliable for $y < 0.8~\mathrm{fm}$. In the region $0.8~\mathrm{fm} < y < 1.1~\mathrm{fm}$, we observe that the factorization ansatz yields the wrong sign. The only approximate agreement with the four-point result can be observed for very large quark distances, where the result is close to zero. Hence, we conclude that within a wide range of $y$, the factorization ansatz is either not reliable because of the mismatch of the support regions, or it fails.

Let us finally discuss the flavor relation \eqref{eq:A_iso}, which follows from our factorization formula but is more general. We see in \fig\ref{fig:uddufact} that the left- and right-hand sides of this relation typically differ by a factor around $2$ in the full $y$ range of the plot. The relation is thus satisfied somewhat better than the factorization hypothesis for $A_{(ud)(du)}$ in terms of three-point correlation functions.

\section{Conclusions}
\label{sec:conclusions}

We extended our work on nucleon DPDs \cite{Bali:2021gel} by considering flavor interference distributions. We calculated the corresponding Wick contractions and extracted the twist-two functions for the flavor combination $(ud)(du)$ for all relevant quark polarizations. This we did for zero proton momentum, which implies $py = 0$. We find that the resulting signal for flavor interference is non-negligible and of the same order of magnitude as the $dd$ contributions. This suggests that at large parton momentum fractions, flavor interference DPDs may not be negligible.

Several relations between the different flavor and polarization combinations can be derived in a simple $SU(6)$ quark model, which can then be compared to the corresponding lattice quantities. We observe good agreement between the lattice data and the $SU(6)$ results for unpolarized quarks if we only take into account the contribution of the $C_1$ contraction, which is most consistent with the picture of a three-quark wave function. If one considers the full connected contribution ($C_1 + C_2$) there is a significant mismatch, which increases for smaller quark distances. The $SU(6)$ model completely fails in the case of two-parton matrix elements for polarized quarks. Finally, we considered the factorization of flavor interference DPDs in terms of GPDs. We found that, depending on the considered quark distance, the factorization ansatz is either not reliable or it does not work in the region where the signal is clearly non-zero.


\section*{Acknowledgments}

This work was in part supported by the Deutsche Forschungsgemeinschaft (DFG, German Research Foundation) by project SCHA 458/23 and FOR 2926, grant number 40965613. The project leading to this publication received funding from the Excellence Initiative of Aix-Marseille University - A*MIDEX, a French “Investissements d’Avenir” programme, AMX-18-ACE-005. We gratefully acknowledge helpful discussions with Gunnar Bali. Moreover, we acknowledge the CLS effort for generating the $n_f = 2 + 1$ ensembles, one of which was used for this work.




\FloatBarrier

\phantomsection

\bibliographystyle{JHEP}
\bibliography{interference}

\providecommand{\href}[2]{#2}\begingroup\raggedright\begin{thebibliography}{10}

\bibitem{Landshoff:1978fq}
P.V.~Landshoff and J.C.~Polkinghorne, \emph{{Calorimeter Triggers for Hard
  Collisions}}, \href{https://doi.org/10.1103/PhysRevD.18.3344}{\emph{Phys.
  Rev. D} {\bfseries 18} (1978) 3344}.

\bibitem{Kirschner:1979im}
R.~Kirschner, \emph{{Generalized {Lipatov-Altarelli-Parisi} Equations and Jet
  Calculus Rules}},
  \href{https://doi.org/10.1016/0370-2693(79)90300-9}{\emph{Phys. Lett. B}
  {\bfseries 84} (1979) 266}.

\bibitem{Politzer:1980me}
H.D.~Politzer, \emph{{Power Corrections at Short Distances}},
  \href{https://doi.org/10.1016/0550-3213(80)90172-8}{\emph{Nucl. Phys. B}
  {\bfseries 172} (1980) 349}.

\bibitem{Paver:1982yp}
N.~Paver and D.~Treleani, \emph{{Multi - Quark Scattering and Large $p_T$ Jet
  Production in Hadronic Collisions}},
  \href{https://doi.org/10.1007/BF02814035}{\emph{Nuovo Cim. A} {\bfseries 70}
  (1982) 215}.

\bibitem{Shelest:1982dg}
V.P.~Shelest, A.M.~Snigirev and G.M.~Zinovev, \emph{{The Multiparton
  Distribution Equations in {QCD}}},
  \href{https://doi.org/10.1016/0370-2693(82)90049-1}{\emph{Phys. Lett. B}
  {\bfseries 113} (1982) 325}.

\bibitem{Mekhfi:1983az}
M.~Mekhfi, \emph{{Multiparton Processes: An Application to Double Drell-Yan}},
  \href{https://doi.org/10.1103/PhysRevD.32.2371}{\emph{Phys. Rev. D}
  {\bfseries 32} (1985) 2371}.

\bibitem{Sjostrand:1986ep}
T.~Sjostrand and M.~van Zijl, \emph{{Multiple Parton-parton Interactions in an
  Impact Parameter Picture}},
  \href{https://doi.org/10.1016/0370-2693(87)90722-2}{\emph{Phys. Lett. B}
  {\bfseries 188} (1987) 149}.

\bibitem{Blok:2010ge}
B.~Blok, Y.~Dokshitzer, L.~Frankfurt and M.~Strikman, \emph{{The Four jet
  production at LHC and Tevatron in QCD}},
  \href{https://doi.org/10.1103/PhysRevD.83.071501}{\emph{Phys. Rev. D}
  {\bfseries 83} (2011) 071501}
  [\href{https://arxiv.org/abs/1009.2714}{{\ttfamily 1009.2714}}].

\bibitem{Gaunt:2011xd}
J.R.~Gaunt and W.J.~Stirling, \emph{{Double Parton Scattering Singularity in
  One-Loop Integrals}},
  \href{https://doi.org/10.1007/JHEP06(2011)048}{\emph{JHEP} {\bfseries 06}
  (2011) 048} [\href{https://arxiv.org/abs/1103.1888}{{\ttfamily 1103.1888}}].

\bibitem{Ryskin:2011kk}
M.G.~Ryskin and A.M.~Snigirev, \emph{{A Fresh look at double parton
  scattering}}, \href{https://doi.org/10.1103/PhysRevD.83.114047}{\emph{Phys.
  Rev. D} {\bfseries 83} (2011) 114047}
  [\href{https://arxiv.org/abs/1103.3495}{{\ttfamily 1103.3495}}].

\bibitem{Blok:2011bu}
B.~Blok, Y.~Dokshitser, L.~Frankfurt and M.~Strikman, \emph{{pQCD physics of
  multiparton interactions}},
  \href{https://doi.org/10.1140/epjc/s10052-012-1963-8}{\emph{Eur. Phys. J. C}
  {\bfseries 72} (2012) 1963}
  [\href{https://arxiv.org/abs/1106.5533}{{\ttfamily 1106.5533}}].

\bibitem{Diehl:2011yj}
{Diehl, Markus and Ostermeier, Daniel and Sch\"afer, Andreas}, \emph{{Elements
  of a theory for multiparton interactions in QCD}},
  \href{https://doi.org/10.1007/JHEP03(2012)089}{\emph{JHEP} {\bfseries 03}
  (2012) 089} [\href{https://arxiv.org/abs/1111.0910}{{\ttfamily 1111.0910}}].

\bibitem{Manohar:2012jr}
A.V.~Manohar and W.J.~Waalewijn, \emph{{A QCD Analysis of Double Parton
  Scattering: Color Correlations, Interference Effects and Evolution}},
  \href{https://doi.org/10.1103/PhysRevD.85.114009}{\emph{Phys. Rev. D}
  {\bfseries 85} (2012) 114009}
  [\href{https://arxiv.org/abs/1202.3794}{{\ttfamily 1202.3794}}].

\bibitem{Manohar:2012pe}
A.V.~Manohar and W.J.~Waalewijn, \emph{{What is Double Parton Scattering?}},
  \href{https://doi.org/10.1016/j.physletb.2012.05.044}{\emph{Phys. Lett. B}
  {\bfseries 713} (2012) 196}
  [\href{https://arxiv.org/abs/1202.5034}{{\ttfamily 1202.5034}}].

\bibitem{Ryskin:2012qx}
M.G.~Ryskin and A.M.~Snigirev, \emph{{Double parton scattering in double
  logarithm approximation of perturbative QCD}},
  \href{https://doi.org/10.1103/PhysRevD.86.014018}{\emph{Phys. Rev. D}
  {\bfseries 86} (2012) 014018}
  [\href{https://arxiv.org/abs/1203.2330}{{\ttfamily 1203.2330}}].

\bibitem{Gaunt:2012dd}
J.R.~Gaunt, \emph{{Single Perturbative Splitting Diagrams in Double Parton
  Scattering}}, \href{https://doi.org/10.1007/JHEP01(2013)042}{\emph{JHEP}
  {\bfseries 01} (2013) 042} [\href{https://arxiv.org/abs/1207.0480}{{\ttfamily
  1207.0480}}].

\bibitem{Blok:2013bpa}
B.~Blok, Y.~Dokshitzer, L.~Frankfurt and M.~Strikman, \emph{{Perturbative QCD
  correlations in multi-parton collisions}},
  \href{https://doi.org/10.1140/epjc/s10052-014-2926-z}{\emph{Eur. Phys. J. C}
  {\bfseries 74} (2014) 2926}
  [\href{https://arxiv.org/abs/1306.3763}{{\ttfamily 1306.3763}}].

\bibitem{Diehl:2017kgu}
M.~Diehl, J.R.~Gaunt and K.~Sch\"onwald, \emph{{Double hard scattering without
  double counting}}, \href{https://doi.org/10.1007/JHEP06(2017)083}{\emph{JHEP}
  {\bfseries 06} (2017) 083}
  [\href{https://arxiv.org/abs/1702.06486}{{\ttfamily 1702.06486}}].

\bibitem{Cabouat:2019gtm}
B.~Cabouat, J.R.~Gaunt and K.~Ostrolenk, \emph{{A Monte-Carlo Simulation of
  Double Parton Scattering}},
  \href{https://doi.org/10.1007/JHEP11(2019)061}{\emph{JHEP} {\bfseries 11}
  (2019) 061} [\href{https://arxiv.org/abs/1906.04669}{{\ttfamily
  1906.04669}}].

\bibitem{Cabouat:2020ssr}
B.~Cabouat and J.R.~Gaunt, \emph{{Combining single and double parton
  scatterings in a parton shower}},
  \href{https://doi.org/10.1007/JHEP10(2020)012}{\emph{JHEP} {\bfseries 10}
  (2020) 012} [\href{https://arxiv.org/abs/2008.01442}{{\ttfamily
  2008.01442}}].

\bibitem{AxialFieldSpectrometer:1986dfj}
{\scshape Axial Field Spectrometer} collaboration, \emph{{Double Parton
  Scattering in $p p$ Collisions at $\sqrt{s}=63$-{GeV}}},
  \href{https://doi.org/10.1007/BF01566757}{\emph{Z. Phys. C} {\bfseries 34}
  (1987) 163}.

\bibitem{CDF:1993sbj}
{\scshape CDF} collaboration, \emph{{Study of four jet events and evidence for
  double parton interactions in $p\bar{p}$ collisions at $\sqrt{s} = 1.8$
  TeV}}, \href{https://doi.org/10.1103/PhysRevD.47.4857}{\emph{Phys. Rev. D}
  {\bfseries 47} (1993) 4857}.

\bibitem{CDF:1997lmq}
{\scshape CDF} collaboration, \emph{{Measurement of double parton scattering in
  $\bar{p}p$ collisions at $\sqrt{s} = 1.8$ TeV}},
  \href{https://doi.org/10.1103/PhysRevLett.79.584}{\emph{Phys. Rev. Lett.}
  {\bfseries 79} (1997) 584}.

\bibitem{D0:2009apj}
{\scshape D0} collaboration, \emph{{Double parton interactions in $\gamma$+3
  jet events in $p p^-$ bar collisions $\sqrt{s}=1.96$ TeV}},
  \href{https://doi.org/10.1103/PhysRevD.81.052012}{\emph{Phys. Rev. D}
  {\bfseries 81} (2010) 052012}
  [\href{https://arxiv.org/abs/0912.5104}{{\ttfamily 0912.5104}}].

\bibitem{D0:2015dyx}
{\scshape D0} collaboration, \emph{{Evidence for simultaneous production of
  $J/\psi$ and $\Upsilon$ mesons}},
  \href{https://doi.org/10.1103/PhysRevLett.116.082002}{\emph{Phys. Rev. Lett.}
  {\bfseries 116} (2016) 082002}
  [\href{https://arxiv.org/abs/1511.02428}{{\ttfamily 1511.02428}}].

\bibitem{LHCb:2012aiv}
{\scshape LHCb} collaboration, \emph{{Observation of double charm production
  involving open charm in pp collisions at $\sqrt{s}$ = 7 TeV}},
  \href{https://doi.org/10.1007/JHEP06(2012)141}{\emph{JHEP} {\bfseries 06}
  (2012) 141} [\href{https://arxiv.org/abs/1205.0975}{{\ttfamily 1205.0975}}].

\bibitem{ATLAS:2013aph}
{\scshape ATLAS} collaboration, \emph{{Measurement of hard double-parton
  interactions in $W(\to l\nu)$+ 2 jet events at $\sqrt{s}$=7 TeV with the
  ATLAS detector}},
  \href{https://doi.org/10.1088/1367-2630/15/3/033038}{\emph{New J. Phys.}
  {\bfseries 15} (2013) 033038}
  [\href{https://arxiv.org/abs/1301.6872}{{\ttfamily 1301.6872}}].

\bibitem{CMS:2013huw}
{\scshape CMS} collaboration, \emph{{Study of Double Parton Scattering Using W
  + 2-Jet Events in Proton-Proton Collisions at $\sqrt{s}$ = 7 TeV}},
  \href{https://doi.org/10.1007/JHEP03(2014)032}{\emph{JHEP} {\bfseries 03}
  (2014) 032} [\href{https://arxiv.org/abs/1312.5729}{{\ttfamily 1312.5729}}].

\bibitem{ATLAS:2016rnd}
{\scshape ATLAS} collaboration, \emph{{Study of hard double-parton scattering
  in four-jet events in pp collisions at $ \sqrt{s}=7 $ TeV with the ATLAS
  experiment}}, \href{https://doi.org/10.1007/JHEP11(2016)110}{\emph{JHEP}
  {\bfseries 11} (2016) 110}
  [\href{https://arxiv.org/abs/1608.01857}{{\ttfamily 1608.01857}}].

\bibitem{LHCb:2015wvu}
{\scshape LHCb} collaboration, \emph{{Production of associated Y and open charm
  hadrons in pp collisions at $ \sqrt{s}=7 $ and 8 TeV via double parton
  scattering}}, \href{https://doi.org/10.1007/JHEP07(2016)052}{\emph{JHEP}
  {\bfseries 07} (2016) 052}
  [\href{https://arxiv.org/abs/1510.05949}{{\ttfamily 1510.05949}}].

\bibitem{LHCb:2020jse}
{\scshape LHCb} collaboration, \emph{{Observation of Enhanced Double Parton
  Scattering in Proton-Lead Collisions at $\sqrt {s_{NN}}$ =8.16 TeV}},
  \href{https://doi.org/10.1103/PhysRevLett.125.212001}{\emph{Phys. Rev. Lett.}
  {\bfseries 125} (2020) 212001}
  [\href{https://arxiv.org/abs/2007.06945}{{\ttfamily 2007.06945}}].

\bibitem{CMS:2022pio}
{\scshape CMS} collaboration, \emph{{Observation of same-sign WW production
  from double parton scattering in proton-proton collisions at $\sqrt{s}$ = 13
  TeV}}, \href{https://doi.org/10.1103/PhysRevLett.131.091803}{\emph{Phys. Rev.
  Lett.} {\bfseries 131} (2023) 091803}
  [\href{https://arxiv.org/abs/2206.02681}{{\ttfamily 2206.02681}}].

\bibitem{Bali:2020mij}
G.S.~Bali, L.~Castagnini, M.~Diehl, J.R.~Gaunt, B.~Gl\"a\ss{}le, A.~Sch\"afer
  et~al., \emph{{Double parton distributions in the pion from lattice QCD}},
  \href{https://doi.org/10.1007/JHEP02(2021)067}{\emph{JHEP} {\bfseries 02}
  (2021) 067} [\href{https://arxiv.org/abs/2006.14826}{{\ttfamily
  2006.14826}}].

\bibitem{Bali:2021gel}
G.S.~Bali, M.~Diehl, B.~Gl\"a\ss{}le, A.~Sch\"afer and C.~Zimmermann,
  \emph{{Double parton distributions in the nucleon from lattice QCD}},
  \href{https://doi.org/10.1007/JHEP09(2021)106}{\emph{JHEP} {\bfseries 09}
  (2021) 106} [\href{https://arxiv.org/abs/2106.03451}{{\ttfamily
  2106.03451}}].

\bibitem{Bruno:2016plf}
M.~Bruno, T.~Korzec and S.~Schaefer, \emph{{Setting the scale for the CLS $2 +
  1$ flavor ensembles}},
  \href{https://doi.org/10.1103/PhysRevD.95.074504}{\emph{Phys. Rev. D}
  {\bfseries 95} (2017) 074504}
  [\href{https://arxiv.org/abs/1608.08900}{{\ttfamily 1608.08900}}].

\bibitem{Bruno:2014jqa}
M.~Bruno et~al., \emph{{Simulation of QCD with N$_{f} =$ 2 $+$ 1 flavors of
  non-perturbatively improved Wilson fermions}},
  \href{https://doi.org/10.1007/JHEP02(2015)043}{\emph{JHEP} {\bfseries 02}
  (2015) 043} [\href{https://arxiv.org/abs/1411.3982}{{\ttfamily 1411.3982}}].

\bibitem{Lin:2017snn}
H.-W.~Lin et~al., \emph{{Parton distributions and lattice QCD calculations: a
  community white paper}},
  \href{https://doi.org/10.1016/j.ppnp.2018.01.007}{\emph{Prog. Part. Nucl.
  Phys.} {\bfseries 100} (2018) 107}
  [\href{https://arxiv.org/abs/1711.07916}{{\ttfamily 1711.07916}}].

\bibitem{Zhang:2023wea}
J.-H.~Zhang, \emph{{Double Parton Distributions from Euclidean Lattice}},
  \href{https://arxiv.org/abs/2304.12481}{{\ttfamily 2304.12481}}.

\bibitem{Jaarsma:2023woo}
M.~Jaarsma, R.~Rahn and W.J.~Waalewijn, \emph{{Towards double parton
  distributions from first principles using Large Momentum Effective Theory}},
  \href{https://doi.org/10.1007/JHEP12(2023)014}{\emph{JHEP} {\bfseries 12}
  (2023) 014} [\href{https://arxiv.org/abs/2305.09716}{{\ttfamily
  2305.09716}}].

\bibitem{Ji:2013dva}
X.~Ji, \emph{{Parton Physics on a Euclidean Lattice}},
  \href{https://doi.org/10.1103/PhysRevLett.110.262002}{\emph{Phys. Rev. Lett.}
  {\bfseries 110} (2013) 262002}
  [\href{https://arxiv.org/abs/1305.1539}{{\ttfamily 1305.1539}}].

\bibitem{Ji:2020ect}
X.~Ji, Y.-S.~Liu, Y.~Liu, J.-H.~Zhang and Y.~Zhao, \emph{{Large-momentum
  effective theory}},
  \href{https://doi.org/10.1103/RevModPhys.93.035005}{\emph{Rev. Mod. Phys.}
  {\bfseries 93} (2021) 035005}
  [\href{https://arxiv.org/abs/2004.03543}{{\ttfamily 2004.03543}}].

\bibitem{Chang:2012nw}
H.-M.~Chang, A.V.~Manohar and W.J.~Waalewijn, \emph{{Double Parton Correlations
  in the Bag Model}},
  \href{https://doi.org/10.1103/PhysRevD.87.034009}{\emph{Phys. Rev. D}
  {\bfseries 87} (2013) 034009}
  [\href{https://arxiv.org/abs/1211.3132}{{\ttfamily 1211.3132}}].

\bibitem{Rinaldi:2013vpa}
M.~Rinaldi, S.~Scopetta and V.~Vento, \emph{{Double parton correlations in
  constituent quark models}},
  \href{https://doi.org/10.1103/PhysRevD.87.114021}{\emph{Phys. Rev. D}
  {\bfseries 87} (2013) 114021}
  [\href{https://arxiv.org/abs/1302.6462}{{\ttfamily 1302.6462}}].

\bibitem{Broniowski:2013xba}
W.~Broniowski and E.~Ruiz~Arriola, \emph{{Valence double parton distributions
  of the nucleon in a simple model}},
  \href{https://doi.org/10.1007/s00601-014-0840-4}{\emph{Few Body Syst.}
  {\bfseries 55} (2014) 381} [\href{https://arxiv.org/abs/1310.8419}{{\ttfamily
  1310.8419}}].

\bibitem{Rinaldi:2014ddl}
M.~Rinaldi, S.~Scopetta, M.~Traini and V.~Vento, \emph{{Double parton
  correlations and constituent quark models: a Light Front approach to the
  valence sector}}, \href{https://doi.org/10.1007/JHEP12(2014)028}{\emph{JHEP}
  {\bfseries 12} (2014) 028} [\href{https://arxiv.org/abs/1409.1500}{{\ttfamily
  1409.1500}}].

\bibitem{Broniowski:2016trx}
W.~Broniowski, E.~Ruiz~Arriola and K.~Golec-Biernat, \emph{{Generalized Valon
  Model for Double Parton Distributions}},
  \href{https://doi.org/10.1007/s00601-016-1087-z}{\emph{Few Body Syst.}
  {\bfseries 57} (2016) 405}
  [\href{https://arxiv.org/abs/1602.00254}{{\ttfamily 1602.00254}}].

\bibitem{Kasemets:2016nio}
T.~Kasemets and A.~Mukherjee, \emph{{Quark-gluon double parton distributions in
  the light-front dressed quark model}},
  \href{https://doi.org/10.1103/PhysRevD.94.074029}{\emph{Phys. Rev. D}
  {\bfseries 94} (2016) 074029}
  [\href{https://arxiv.org/abs/1606.05686}{{\ttfamily 1606.05686}}].

\bibitem{Rinaldi:2016jvu}
M.~Rinaldi, S.~Scopetta, M.C.~Traini and V.~Vento, \emph{{Correlations in
  Double Parton Distributions: Perturbative and Non-Perturbative effects}},
  \href{https://doi.org/10.1007/JHEP10(2016)063}{\emph{JHEP} {\bfseries 10}
  (2016) 063} [\href{https://arxiv.org/abs/1608.02521}{{\ttfamily
  1608.02521}}].

\bibitem{Rinaldi:2016mlk}
M.~Rinaldi and F.A.~Ceccopieri, \emph{{Relativistic effects in model
  calculations of double parton distribution function}},
  \href{https://doi.org/10.1103/PhysRevD.95.034040}{\emph{Phys. Rev. D}
  {\bfseries 95} (2017) 034040}
  [\href{https://arxiv.org/abs/1611.04793}{{\ttfamily 1611.04793}}].

\bibitem{Rinaldi:2018zng}
M.~Rinaldi, S.~Scopetta, M.~Traini and V.~Vento, \emph{{A model calculation of
  double parton distribution functions of the pion}},
  \href{https://doi.org/10.1140/epjc/s10052-018-6256-4}{\emph{Eur. Phys. J. C}
  {\bfseries 78} (2018) 781}
  [\href{https://arxiv.org/abs/1806.10112}{{\ttfamily 1806.10112}}].

\bibitem{Courtoy:2019cxq}
A.~Courtoy, S.~Noguera and S.~Scopetta, \emph{{Double parton distributions in
  the pion in the Nambu\textendash{}Jona-Lasinio model}},
  \href{https://doi.org/10.1007/JHEP12(2019)045}{\emph{JHEP} {\bfseries 12}
  (2019) 045} [\href{https://arxiv.org/abs/1909.09530}{{\ttfamily
  1909.09530}}].

\bibitem{Broniowski:2019rmu}
W.~Broniowski and E.~Ruiz~Arriola, \emph{{Double parton distribution of valence
  quarks in the pion in chiral quark models}},
  \href{https://doi.org/10.1103/PhysRevD.101.014019}{\emph{Phys. Rev. D}
  {\bfseries 101} (2020) 014019}
  [\href{https://arxiv.org/abs/1910.03707}{{\ttfamily 1910.03707}}].

\bibitem{Broniowski:2020jwk}
W.~Broniowski and E.~Ruiz~Arriola, \emph{{Double parton distributions of the
  pion in the NJL model}},
  \href{https://doi.org/10.22323/1.374.0031}{\emph{PoS} {\bfseries LC2019}
  (2020) 031} [\href{https://arxiv.org/abs/2001.00883}{{\ttfamily
  2001.00883}}].

\bibitem{RQCD:2020kuu}
{\scshape RQCD} collaboration, \emph{{Nonperturbative Renormalization in
  Lattice QCD with three Flavors of Clover Fermions: Using Periodic and Open
  Boundary Conditions}},
  \href{https://doi.org/10.1103/PhysRevD.103.094511}{\emph{Phys. Rev. D}
  {\bfseries 103} (2021) 094511}
  [\href{https://arxiv.org/abs/2012.06284}{{\ttfamily 2012.06284}}].

\bibitem{Bali:2016lva}
G.S.~Bali, B.~Lang, B.U.~Musch and A.~Sch\"afer, \emph{{Novel quark smearing
  for hadrons with high momenta in lattice QCD}},
  \href{https://doi.org/10.1103/PhysRevD.93.094515}{\emph{Phys. Rev. D}
  {\bfseries 93} (2016) 094515}
  [\href{https://arxiv.org/abs/1602.05525}{{\ttfamily 1602.05525}}].

\bibitem{Zimmermann:2019quf}
{\scshape RQCD} collaboration, \emph{{Two-current correlations and DPDs for the
  nucleon on the lattice}},
  \href{https://doi.org/10.22323/1.363.0040}{\emph{PoS} {\bfseries LATTICE2019}
  (2019) 040} [\href{https://arxiv.org/abs/1911.05051}{{\ttfamily
  1911.05051}}].

\bibitem{Burkardt:1994pw}
M.~Burkardt, J.M.~Grandy and J.W.~Negele, \emph{{Calculation and interpretation
  of hadron correlation functions in lattice QCD}},
  \href{https://doi.org/10.1006/aphy.1995.1026}{\emph{Annals Phys.} {\bfseries
  238} (1995) 441} [\href{https://arxiv.org/abs/hep-lat/9406009}{{\ttfamily
  hep-lat/9406009}}].

\bibitem{Alexandrou:2008ru}
C.~Alexandrou and G.~Koutsou, \emph{{A Study of Hadron Deformation in Lattice
  QCD}}, \href{https://doi.org/10.1103/PhysRevD.78.094506}{\emph{Phys. Rev. D}
  {\bfseries 78} (2008) 094506}
  [\href{https://arxiv.org/abs/0809.2056}{{\ttfamily 0809.2056}}].

\bibitem{Bali:2018spj}
G.S.~Bali, V.M.~Braun, B.~Gl\"a\ss{}le, M.~G\"ockeler, M.~Gruber, F.~Hutzler
  et~al., \emph{{Pion distribution amplitude from Euclidean correlation
  functions: Exploring universality and higher-twist effects}},
  \href{https://doi.org/10.1103/PhysRevD.98.094507}{\emph{Phys. Rev. D}
  {\bfseries 98} (2018) 094507}
  [\href{https://arxiv.org/abs/1807.06671}{{\ttfamily 1807.06671}}].

\bibitem{Cichy:2012is}
K.~Cichy, K.~Jansen and P.~Korcyl, \emph{{Non-perturbative renormalization in
  coordinate space for $N_f=2$ maximally twisted mass fermions with tree-level
  Symanzik improved gauge action}},
  \href{https://doi.org/10.1016/j.nuclphysb.2012.08.006}{\emph{Nucl. Phys. B}
  {\bfseries 865} (2012) 268}
  [\href{https://arxiv.org/abs/1207.0628}{{\ttfamily 1207.0628}}].

\bibitem{Mankiewicz:1997aa}
L.~Mankiewicz, G.~Piller and T.~Weigl, \emph{{Hard leptoproduction of charged
  vector mesons}},
  \href{https://doi.org/10.1103/PhysRevD.59.017501}{\emph{Phys. Rev. D}
  {\bfseries 59} (1999) 017501}
  [\href{https://arxiv.org/abs/hep-ph/9712508}{{\ttfamily hep-ph/9712508}}].

\end{thebibliography}\endgroup

\end{document}